\documentclass{aa}
\usepackage{graphicx}
\usepackage{txfonts}
\usepackage{longtable}
\IfFileExists{url.sty}{\usepackage{url}}
                      {\newcommand{\url}{\texttt}}
\def\Ep{$E_{peak}$}
\def\Epobs{$E_{peak}^{obs}$}
\def\Epi{$E_{peak}^{intr}$}
\def\T90{$T_{90}^{obs}$}
\def\T90i{$T_{90}^{intr}$}
\def\Eiso{$E_{iso}$}
\def\zsamp{$z_{sample}$}
\begin{document}
\title{Intrinsic properties of a complete sample\\ 
of  {\em HETE-2} Gamma-ray bursts}
\subtitle{A measure of the GRB rate in the Local Universe}
\author{
A.~P\'elangeon\inst{1}
\and J-L.~Atteia\inst{1}
\and Y.~E.~Nakagawa\inst{2,3}
\and K.~Hurley\inst{5}
\and A.~Yoshida\inst{2,3}
\and R.~Vanderspek\inst{4}
\and M.~Suzuki\inst{14}
\and N.~Kawai\inst{3,6}
\and G.~Pizzichini\inst{7}
\and M.~Bo\"er\inst{8}
\and J.~Braga\inst{9}
\and G.~Crew\inst{4}
\and T.~Q.~Donaghy\inst{10}
\and J.~P.~Dezalay\inst{11}
\and J.~Doty\inst{4}
\and E.~E.~Fenimore\inst{12}
\and M.~Galassi\inst{12}
\and C.~Graziani\inst{10}
\and J.~G.~Jernigan\inst{5}
\and D.~Q.~Lamb\inst{10}
\and A.~Levine\inst{4}
\and J.~Manchanda\inst{13}
\and F.~Martel\inst{4}
\and M.~Matsuoka\inst{14}
\and J-F.~Olive\inst{11}
\and G.~Prigozhin\inst{4}
\and G.~R.~Ricker\inst{4}
\and T.~Sakamoto\inst{15,20,21}
\and Y.~Shirasaki\inst{16} 
\and S.~Sugita\inst{17}
\and K.~Takagishi\inst{18}
\and T.~Tamagawa\inst{3}
\and J.~Villasenor\inst{4}
\and S.~E.~Woosley\inst{19}
\and M.~Yamauchi\inst{18}
}
\institute{
Laboratoire d'Astrophysique de Toulouse-Tarbes, 
Universit\'e de Toulouse, CNRS, 
14 Av Edouard Belin, 31400 Toulouse, France
\and Department of Physics and Mathematics, 
Aoyama Gakuin University, 5-10-1 Fuchinobe, 
Sagamihara, Kanagawa 229-8558, Japan
\and RIKEN (Institute of Physical and Chemical Research), 
2-1 Hirosawa, Wako, Saitama 351-0198, Japan
\and Center for Space Research, Massachusetts Institute of Technology, 
70 Vassar Street, Cambridge, MA, 02139, USA
\and Space Sciences Laboratory, University of California 
at Berkeley, 7 Gauss Way, Berkeley, CA, 94720-7450, USA
\and Department of Physics, Tokyo Institute of Technology, 
2-12-1 Ookayama, Meguro-ku, Tokyo, 152-8551, Japan
\and INAF/IASF Bologna, via Gobetti 101, 40129 Bologna, Italy
\and Observatoire de Haute-Provence (CNRS/OAMP), 
Saint Michel l'Observatoire, France
\and Instituto Nacional de Pesquisas Espaciais, 
Avenida Dos Astraunotas 1758, S\~ao Jose dos Campos 
12227-010, Brazil
\and Department of Astronomy and Astrophysics, 
University of Chicago, 5640 South Ellis Avenue, 
Chicago, IL, 60637, USA
\and Centre d'Etude Spatiale des Rayonnements, 
Universit\'e de Toulouse, CNRS, 
9 Av du Colonel Roche, 31028 Toulouse, France
\and Los Alamos National Laboratory, P.O.~Box 1663, 
Los Alamos, NM, 87545, USA
\and Department of Astronomy and Astrophysics, 
Tata Institute of Fundamental Research, 
Homi Bhabha Road, Mumbai, 400~005, India
\and Tsukuba Space Center, National Space Development 
Agency of Japan, Tsukuba, Ibaraki, 305-8505, Japan
\and NASA Goddard Space Flight Center, Greenbelt, MD, 20771, USA
\and National Astronomical Observatory, 2-21-1, Osawa, Mitaka, 
Tokyo, 181-8588, Japan
\and Department of Physics, Aoyama Gakuin University, 
Chitosedai, 6-16-1, Setagaya-ku, Tokyo, 157-8572, Japan
\and Faculty of Engineering, Miyazaki University, 
Gakuen Kibanadai Nishi, Miyazaki, 889-2192, Japan
\and Department of Astronomy and Astrophysics, 
Univ of California at Santa Cruz, 
477 Clark Kerr Hall, Santa Cruz, CA, 95064, USA 
\and Center for Research and Exploration in Space Science 
and Technology (CRESST), NASA Goddard Space Flight Center, 
Greenbelt, MD 20771, USA
\and Joint Center for Astrophysics, University of Maryland, 
Baltimore County, 1000 Hilltop Circle, Baltimore, MD 21250, USA
}
\offprints{A. P\'elangeon \& J-L. Atteia \\
\email{[alexandre.pelangeon/atteia]@ast.obs-mip.fr}}
\date{Received / Accepted }
\titlerunning{Intrinsic properties of a complete sample of {\em HETE-2} GRBs}
\authorrunning{P\'elangeon {\it et al.}}
%
%
%
\abstract
{As a result of the numerous missions dedicated to the detection of Gamma-ray bursts (GRBs), 
the observed properties of these events are now well known. 
However, studying their parameters in the source frame is not simple since it requires 
having measurements of both the bursts' parameters and of their distances.} 
{Taking advantage of the forthcoming Catalog of the High Energy Transient Explorer 2
(HETE-2) mission, the aim of this paper is to evaluate the main properties of HETE-2 GRBs -- the 
peak energy (\Ep), the duration ($T_{90}$) and the isotropic energy (\Eiso) -- in their source 
frames and to derive their unbiased distribution.}
{We first construct a complete sample containing all the bursts localized by the Wide-Field 
X-ray Monitor (WXM) on-board HETE-2, which are selected with a uniform criterion and whose 
observed parameters can be constrained. We then derive the intrinsic \Ep, $T_{90}$ and \Eiso\ 
distributions using their redshift when it is available, or their pseudo-redshift otherwise.
We finally compute the `volume of detectability' $V_{max}$ of each GRB, i.e. the volume of the 
universe in which the burst is bright enough to be part of our sample, and the corresponding 
number of GRB within their visibility volume $N_{Vmax}$, in order to derive a weight for each 
detected burst accounting both for the detection significance and the star formation history 
of the universe.}
{We obtain unbiased distributions of three intrinsic properties of HETE-2 GRBs: \Epi, 
$T_{90}^{intr}$ and the isotropic energy of the burst. These distributions clearly show the 
predominence of X-ray flashes (XRFs) in the global GRB population. We also derive the rate of 
local GRBs: $R_{0}^\mathrm{H2}\gtrsim$~11~Gpc$^{-3}$yr$^{-1}$, which is intermediate between 
the local rate obtained by considering only the `high-luminosity' bursts 
($\sim$1~Gpc$^{-3}$yr$^{-1}$) and that obtained by including the `low-luminosity' bursts 
($\gtrsim$~200~Gpc$^{-3}$yr$^{-1}$).}
{This study shows that the XRFs are predominent in the GRB population and are closely linked to 
the `classical' GRBs. We show that HETE-2 detected no low-luminosity GRB like GRB~980425 or 
XRF~060218, due to the small size of its detectors, excluding this type of burst from our 
statistical analysis. The comparison of the GRB rate derived in this study with the known rate
of Type Ib/c supernovae clearly shows that the progenitors of SNe~Ib/c must have some special 
characteristics in order to produce a gamma-ray burst or an X-ray flash.\\
\\}
\maketitle
%
%
%
\section{Introduction}
\label{intro}
In recent years we have learned that long Gamma-ray bursts ($T_{90}$~$>$~2~sec.) 
are associated with the death of massive stars ($M\gtrsim$~20-30~$M_{\sun}$).
This origin has been clearly established by the association of a few nearby GRBs
with Type Ib/c supernovae (SNe~Ib/c): GRB~980425 detected by Beppo-SAX (Pian et al. \cite{pian99}) 
and SN~1998bw (Galama et al. \cite{galama98}; Kulkarni et al. \cite{kulkarni98}), 
GRB~021211 detected by HETE-2 (Crew et al. \cite{crew02}, \cite{crew03}) 
and SN~2002lt (Della Valle et al. \cite{dellavalle03}), GRB~030329 also detected 
by HETE-2 (Vanderspek et al. \cite{vanderspek03}, \cite{vanderspek04}) 
and SN~2003dh (e.g. Matheson et al. \cite{matheson03}; Stanek et al. \cite{stanek03}; 
Hjorth et al. \cite{hjorth03a}; Mazzali et al. \cite{mazzali03}), 
GRB~031203 detected by INTEGRAL (G\"otz et al. \cite{gotz03}) and SN~2003lw 
(Malesani et al. \cite{malesani04}), GRB~050525A detected by Swift-BAT (Band et al. \cite{band05}) 
and SN~2005nc (Della Valle et al. \cite{dellavalle06b}). The association of the nearby GRB~060218 
detected by Swift-BAT (Cusumano et al. \cite{cusumano06}) with SN~2006aj 
(Masetti et al. \cite{masetti06}; Modjaz et al. \cite{modjaz06};
Pian et al. \cite{pian06}; Campana et al. \cite{campana06}; 
Sollerman et al. \cite{sollerman06}; Mirabal et al. \cite{mirabal06b}; Cobb et al. \cite{cobb06}) 
provided a remarkable example of this paradigm.
Even though a massive star's explosion can explain the production of both the GRB and the supernova
(see for instance the collapsar model proposed by Woosley \cite{woosley93}; 
see also Woosley \& Bloom \cite{woosley06}; Della Valle \cite{dellavalle06a}), these two
phenomena are profoundly different in nature: the GRB is due to an ultra-relativistic
outflow generated by a newborn black hole or magnetar, while the supernova is powered
by the radioactive decay of $^{56}$Ni within a massive shell of matter ejected 
at sub-relativistic velocities (e.g. Soderberg et al. \cite{soderberg06d}). 
The question of the link between GRBs and SNe has recently become more complicated 
by the discovery of two low-redshift, long gamma-ray bursts -- GRB~060505 at $z=$0.089 and
GRB~060614 at $z=$0.125 -- which were not associated with a supernova
(Gehrels et al. \cite{gehrels06}; Fynbo et al. \cite{fynbo06b}; Della Valle et al \cite{dellavalle06d};
Gal-Yam et al. \cite{gal-yam06}). It is thus fair to say that despite very significant advances in our
understanding of GRBs, the global picture continues to escape us and key issues remain
to be elucidated. Concerning the connection between GRBs and supernovae, the observational clues 
are based on two facts: the clear association of some GRBs with Type Ib/c supernovae,
and the evidence, based on statistical studies, that the majority
of the 9000 Type Ib/c supernovae exploding each year in the universe do not produce GRBs.
This last point was addressed by Berger et al. (\cite{berger03}),
Soderberg et al. (\cite{soderberg06a}, \cite{soderberg06b}, \cite{soderberg06c}, \cite{soderberg06d}, 
\cite{soderberg04b}) and Della Valle (\cite{dellavalle05}, \cite{dellavalle06a}, \cite{dellavalle06c}).\\
\\
While the SNe~Ib/c population is rather well sampled in our local
environment, the situation is quite different for GRBs, which need a complex 
series of detections to be fully characterized: detection of the high-energy
signal in space, quick transmission to the ground, identification of the
afterglow at X-ray, optical or radio wavelengths, and measurement of the redshift.
This situation makes our understanding of the GRB population much less secure
than for supernovae. Indeed, the observations by Beppo-SAX, HETE-2, and now Swift 
have demonstrated the diversity of the GRB population, which is composed of classical GRBs, 
short-duration GRBs, X-ray flashes (XRFs), and low-luminosity GRBs.
In this classification, the `classical GRBs' appear to be the `high-luminosity' part (HL-GRBs) 
of a population which could consist mostly of low-luminosity or subluminous bursts (LL-GRBs),
like GRB~980425 (Tinney at al. \cite{tinney98}) and GRB~060218 (Mirabal et al \cite{mirabal06a}), 
or of X-ray flashes.\\
\\
In this paper we take advantage of the broad energy coverage of HETE-2 
instruments (2-400~keV) to discuss the global properties of the GRB population, 
and to infer a rate of GRB+XRF events which can be compared with
theoretical predictions and with the known rate of Type Ib/c supernovae.
This work is partly based on the forthcoming catalog of HETE-2 GRBs
(Vanderspek et al. \cite{vanderspek08}). For a description of the HETE-2 mission 
and its instrumentation, see Ricker et al. (\cite{ricker01}), Atteia et al. (\cite{atteia03a}) 
and Villasenor et al. (\cite{villasenor03}).\\
Our work relies on the construction of a complete GRB sample 
containing all the long-duration bursts localized by the {\em Wide Field X-Ray Monitor (WXM)} 
and having  a signal-to-noise ratio (SNR) larger than a given threshold in {\em FREGATE} 
({\em FREnch GAmma-ray TElescope}) or in the WXM. For each burst in this sample, 
we compute the intrinsic properties at the source ($T_{90}$, \Ep, and \Eiso) by correcting 
the observed properties for the effect of the redshift.
When a spectroscopic redshift is not available (62 GRBs out of 82), 
we use the {\em pseudo-redshift} following the method of P\'elangeon et al., 
i.e. the estimate of the redshift which is partly based 
on both the \Ep--\Eiso\ (e.g. Amati et al. \cite{amati02}) 
and \Ep--$L_{iso}$ (Yonetoku et al. \cite{yonetoku04}) correlations, 
and derived from the spectral properties of the prompt emission 
(for more details see P\'elangeon et al. \cite{pelangeon06a} 
and P\'elangeon \& Atteia \cite{pelangeon06b}).
We emphasize that this procedure has little impact on the
final results since it introduces an additional uncertainty which is
much smaller than the intrinsic dispersion of the parameters under study.
In a second step we derive the `visibility distance' for each burst 
($z_{max}$), the distance at which the SNR of the burst reaches the threshold
of our analysis, and the number of GRB within its visibility volume ($N_{Vmax}$).
We then attribute to each GRB a weight $W=1/N_{Vmax}$, accounting both 
for the detection significance and the star formation history of the universe. 
Consequently, this method enables us to renormalize the global distribution, 
taking into account the true rate of occurrence of each type of GRB. 
We finally derive the GRB rate detected by HETE-2, and we discuss 
the relative importance of XRFs and classical GRBs in the overall GRB population.\\
\\
This paper is organized as follows: Section~\ref{s2} is devoted to the description
of the HETE-2 GRB sample. Section~\ref{s3} references and describes the tools necessary for
our study. The three parameters studied here, \Ep, $T_{90}$, and \Eiso, 
are discussed in Sections~\ref{s4}, \ref{s5} and \ref{s6} respectively.
The remaining three sections are dedicated to the interpretation of the results.
In Section~\ref{s7}, we check the existence of correlations between the intrinsic parameters.
In Section~\ref{s8}, we show that our study gives some clues to the nature of X-ray flashes
in terms of intrinsic energetics and distance-scales. Finally,
in Section~\ref{s9} we discuss the rate of GRBs in the local universe and compare it with
previous estimates obtained by other authors. A summary of the main results 
obtained in this paper is presented in Section~\ref{s10}.
%
%
%
\section{Obtaining of a complete sample of HETE-2 bursts}
\label{s2}
\subsection{Instrumentation}
Since its launch in October 2000, the HETE-2 satellite (Ricker et al. \cite{ricker01}), 
dedicated to the detection and observation of Gamma-Ray Bursts, 
has detected 250 events classified as {\it GRBs} 
(Vanderspek et al. \cite{vanderspek08}). 
Thanks to the combination of its 3 instruments, the {\em FREnch GAmma-ray TElescope} 
(FREGATE, Atteia et al. \cite{atteia03a}), the {\em Wide-field X-ray Camera} 
(WXM, Shirasaki et al. \cite{shirasaki03}) and the {\em Soft X-ray Camera} 
(SXC, Villasenor et al. \cite{villasenor03}), HETE-2 has a broad energy range 
covering the hard $X$-rays and $\gamma$-rays (2-400~keV), allowing the detection of both 
{\em classical} and {\em soft} GRBs (XRFs). Moreover, HETE-2 was designed to provide GRB 
positions to the community through the {\em GRB Coordinates Network (GCN)} 
(Barthelmy et al. \cite{barthelmy00}) with an accuracy of one to several arcminutes, 
within seconds of the trigger. Consequently, multi-wavelength ground follow-up 
was done for most HETE-2 GRBs, leading to the determination of 25 spectroscopic redshifts.
\subsection{Burst selection} 
It is of prime importance for our study to construct 
a GRB sample for which the detection criteria are fully understood
and can be reproduced. This is essential for the determination
of the visibility volume of each GRB in the sample. In addition
it is necessary to be able to compute for each burst the three parameters
of our study: \Ep, $T_{90}$, and the fluence.
This section explains the construction of 
the sample\footnote{The order of the steps described in this Section 
is not important; we could have performed them in any order and 
obtained the same sample.}.
\subsubsection{Availability of the spectral data}
We considered all the 250 events that constitute the total sample 
detected by HETE-2, i.e. both triggered and untriggered bursts, 
thus reducing the bias due to trigger algorithms. Then, we selected 
the bursts for which the angle of incidence was measured by any experiment. 
Its knowledge is necessary to perform the spectral analysis of the bursts 
and to derive their \Ep\ and \Eiso. Of the 250 GRBs, 132 have 
no accurate localization and were rejected, leaving 118 GRBs.\\
Among these bursts, we have also rejected GRB~040810 because this long and intense burst 
occurred as the instruments were being shut down. Consequently only the 
precursor was detected by HETE-2.
\subsubsection{Technical problems} 
Of the remaining 117 GRBs, 2 bursts were rejected due to technical problems: 
GRB~010110 was detected and localized by the WXM when FREGATE 
was not working. It is thus impossible to constrain 
the spectral parameters with only the WXM data. The second burst rejected 
is GRB~021113. At the moment of its detection there were onboard software 
problems that did not permit the trigger to reach the ground. 
The spectral data are thus not available.
\subsubsection{WXM localization} 
\begin{figure}[ht]
\begin{center}
\includegraphics[width=1\linewidth,angle=0]{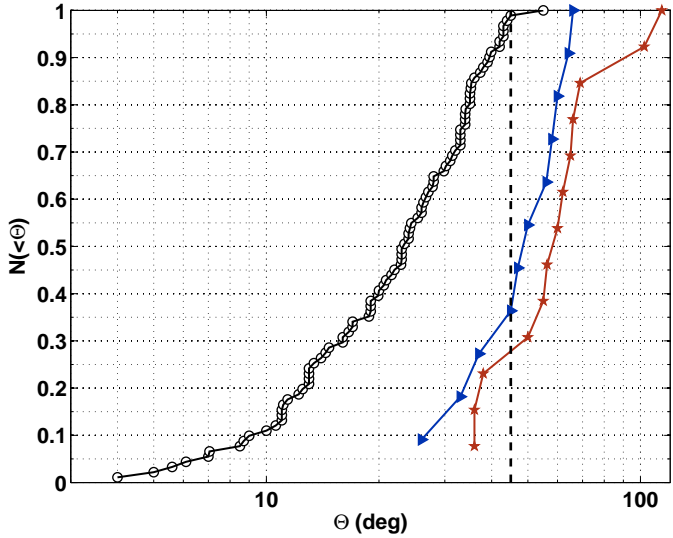}
\end{center} 
\caption{\label{selec_angle} Boresight angle cumulative distribution of the 115 GRBs localized 
by HETE-2/WXM, the Interplanetary Network or Swift. The dark dashed vertical line 
represents the `angle limit' of 45$\degr$, corresponding to the limit 
of detection by the WXM. The GRBs localized by the WXM are represented 
with black open circles, the ones localized by the IPN are in blue filled triangles, 
and the ones localized by Swift are in dark-red filled stars.}
\end{figure}

Considering the 115 remaining GRBs with a known angle, 
we removed 24 GRBs for which the angle was obtained through the localization 
by the {\it Inter Planetary Network} (IPN)
or the {\em Swift} satellite (Gehrels et al. \cite{gehrels04}). 
Indeed, as the IPN and Swift have different sensitivities, we would 
have added a bias to our sample by considering them.\\ 
Moreover, in order to have bursts well within the field of view of the WXM, 
we have considered an `incident angle limit' of 45$\degr$. 
This angle corresponds to a limit of detection for the WXM 
(Shirasaki et al. \cite{shirasaki03}), and to a decrease 
of the effective area for FREGATE (Atteia et al. \cite{atteia03a}). 
With this cut, only one burst is rejected: GRB~020201, whose angle is 55$\degr$. 
This step leaves 90 GRBs in the sample (see Fig.~\ref{selec_angle}).
\subsubsection{Short-duration bursts}
We also rejected the two bursts classified without ambiguity 
as short-duration bursts: GRB~020531 (Lamb et al. \cite{lamb03}) 
and GRB~050709 (Villasenor et al. \cite{villasenor05}).
This is based on the fact that this class of burst is probably not 
associated with the same progenitors as the long GRBs. As one goal 
of this paper is to discuss the rate of long GRBs in the local universe and to compare 
it with the rate of SNe, we have excluded the short-duration bursts, 
leaving 88 bursts in the sample.
\subsubsection{Threshold cut}
Finally, in order to have homogeneous detection criteria for all the events 
of the sample, we removed the GRBs with a signal-to-noise ratio (SNR) 
lower than a given threshold in FREGATE and in the WXM. 
These SNR were computed as follows: 
\begin{itemize}
\item we measured the peak count rate of each burst. For FREGATE we used 
two different time resolutions (1.3 and 5.2~s) and two different energy ranges 
(6-80 and 30-400~keV) corresponding to the bands used by the instrument to trigger. 
For the WXM we used three different time resolutions (1.2, 4.9 and 9.8~s) 
and the total energy range of the WXM (2-25~keV). 
\item we computed a SNR for each burst in all the combinations 
of time resolution and energy range. We kept only the GRBs for which the SNR 
exceeded 7.9 in FREGATE or 5.8 in the WXM (in at least one combination 
of the time and energy ranges, see Fig. \ref{selec_snr}).
\end{itemize}
This last criterion removed 6 bursts with a low SNR: 
GRB~011103, XRF~020903, GRB~030323, GRB~030706, GRB~040131 and GRB~040228.\\
\\
Our final sample contains 82 GRBs whose properties 
are given in Tables \ref{tab1} and \ref{tab2}.
\subsection{The cases of XRF~020903 and GRB~030323} 
\begin{figure}[ht]
\begin{center}
\includegraphics[width=1\linewidth,angle=0]{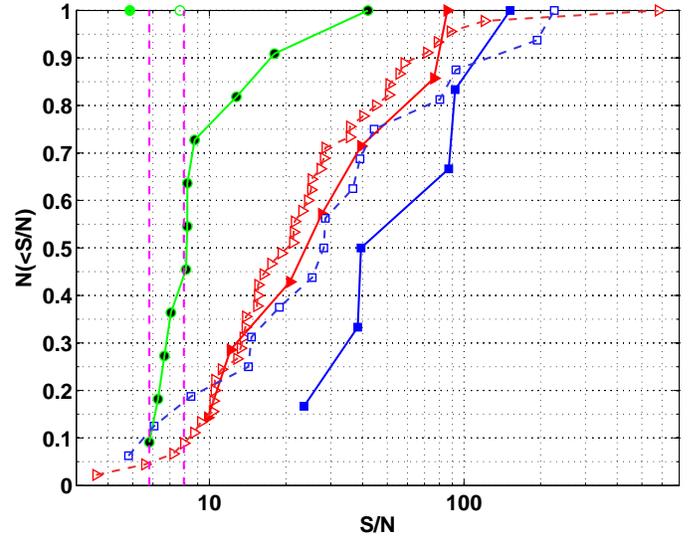}
\end{center} 
\caption{\label{selec_snr} Signal-to-noise ratio (SNR) cumulative distribution 
of the 88 GRBs localized by the WXM and within 45$\degr$ of the boresight. 
The threshold of FREGATE (SNR$=$7.9), is represented with the right-most vertical magenta 
dashed line, whereas the threshold of the WXM (SNR$=$5.8) is represented 
with the left-most vertical magenta dashed line. The different symbols show 
the configuration of time resolution and energy range leading to the highest 
SNR for each burst: 
a given shape is associated with a given energy range 
and a given color corresponds to a given time resolution:
[WXM/2-25~keV/1.2~s]= light-green filled circles; 
[WXM/2-25~keV/4.9~s]= light-green open circles;
[WXM/2-25~keV/9.8~s]= dark filled circles; 
[FREG/6-80~keV/1.3~s]= red filled triangles; 
[FREG/6-80~keV/5.2~s]= red open triangles;
[FREG/30-400~keV/1.3~s]= blue filled squares; 
[FREG/30-400~keV/5.2~s]= blue open squares. Note that there is only one 
light-green filled circle and one light-green open circle (left part of the 
figure) due to the fact that only one burst has its highest SNR in the 
configuration [WXM/2-25~keV/1.2~s] and one burst in the configuration 
[WXM/2-25~keV/4.9~s].}
\end{figure}

These two bursts have a spectroscopic redshift but they are rejected 
from our sample due to their SNR 
just below the thresholds of FREGATE and the WXM (Fig.~\ref{selec_snr}). 
As they have spectroscopic redshifts -- $z=0.25$ for XRF~020903 
(Soderberg et al. \cite{soderberg04a}) and $z=3.372$ for GRB~030323 
(Vreeswijk et al. \cite{vreeswijk03}) -- we can argue that even if they had been 
present in our sample, the results presented in this paper would actually 
have been reinforced. Indeed, situated at a moderately high redshift, GRB~030323 
has a large visibility volume and thus a low weight. On the other hand, 
the nearby XRF~020903 has a small visibility volume 
and hence a large weight. Since this is an XRF with a low \Ep, 
adding it to our GRB sample would have increased the predominance of X-ray flashes 
which is shown in Section~\ref{s4}, and strengthened the conclusions of this paper.
%
%
%
\section{Deriving the intrinsic properties of HETE-2 GRBs}
\label{s3}
\subsection{Burst parameters}
\label{cat}
We have used the spectral parameters and durations available 
in the HETE-2 Catalog (Vanderspek et al. \cite{vanderspek08}) 
which summarizes the information on GRBs detected by HETE-2 during 
the entire mission and thus complements the previous spectral analyses 
based on smaller samples of HETE-2 GRBs performed by Barraud et al. 
(\cite{barraud03}, \cite{barraud04}) and Sakamoto et al. (\cite{sakamoto05}). 
\subsection{Burst distance}
At the end of the 90s, the growing sample of GRBs with measured redshift 
allowed new types of studies: searches for correlations between physical 
quantities -- characterizing the gamma-ray bursts' light-curves 
and/or spectra -- {\em in their rest frame}. 
Following the discovery of correlations between \Eiso, the isotropic 
equivalent energy emitted by the burst, and various intrinsic GRB properties, 
it has been suggested that some GRB observables could be used as luminosity indicators, 
and hence as redshift indicators. The most used are the {\it lag-luminosity} 
($\tau_{lag}$--$L_{p,iso}$) correlation (Norris, Marani \& Bonnell \cite{norris00}), 
the {\it variability-luminosity} ($V$--$L_{p,iso}$) correlation  (e.g. Fenimore \& Ramirez-Ruiz \cite{fenimore00}; 
Reichart \& al. \cite{reichart01}; Lloyd-Ronning, Fryer \& Ramirez-Ruiz \cite{lloyd-ronning02a}) 
and the \Ep--\Eiso\ correlation -- suggested by several authors (see e.g. Lloyd, 
Petrosian \& Mallozzi \cite{lloyd00}; Atteia \cite{atteia00}; 
Lloyd-Ronning \& Ramirez-Ruiz \cite{lloyd-ronning02b}) -- before Amati et al. 
(\cite{amati02}) firmly established it with a study of 12 GRBs with known redshift. 
Other correlations were found later, such as the \Ep--$L_{p,iso}$ correlation 
(Yonetoku et al. \cite{yonetoku04}), the \Ep--$E_{\gamma}$ correlation 
(Ghirlanda, Ghisellini \& Lazzati \cite{ghirlanda04}) 
and the \Ep--$L_{p,iso}$--$T_{0.45}$ correlation (Firmani et al. \cite{firmani06}). 
Recently, in his construction of the Hubble Diagram from a sample 
of 69 GRBs with redshifts, Schaefer (\cite{schaefer07}) showed that 
the use of five of these correlations simultaneously 
-- lag-luminosity ($\tau_{lag}$--$L_{p,iso}$, Norris et al. \cite{norris00}); 
$V$--$L_{p,iso}$ (Fenimore \& Ramirez-Ruiz \cite{fenimore00}); 
\Ep--$L_{p,iso}$ (Schaefer \cite{schaefer03}); \Ep--$E_{\gamma}$ (Ghirlanda et al. \cite{ghirlanda04}) 
and the minimum light-curves rise time versus luminosity 
($\tau_{RT}-L_{p,iso}$, Schaefer \cite{schaefer02})\footnote{See e.g. Ghirlanda, 
Ghisellini \& Firmani (\cite{ghirlanda06}) or Schaefer (\cite{schaefer07}) 
for a complete summary of these correlations.} -- led to more 
reliable luminosity distances than the use of only one distance indicator.\\
The design of HETE-2 permitted the study of the energetics 
at work in GRBs, both in the observer and in the source frames. For example, 
the \Ep-\Eiso\ correlation (Amati et al. \cite{amati02}) 
was confirmed and extended at lower energy, thanks to the sample 
of HETE-2 X-ray rich GRBs (XRRs) and X-ray flashes (Amati \cite{amati03}; 
Lamb, Donaghy \& Graziani \cite{lamb04a}; Lamb et al. \cite{lamb04b}; 
Amati \cite{amati06}), particularly XRF~020903 (Sakamoto et al. \cite{sakamoto04}). 
The HETE-2 sample is thus appropriate 
for the determination of {\em pseudo-redshifts} based on the \Ep-\Eiso\ correlation. 
A first attempt in this direction was proposed by Atteia (\cite{atteia03b}), 
which was later revised by P\'elangeon et al. (\cite{pelangeon06a}) to take 
into account the complex lightcurves of some long GRBs\footnote{See also \url{http://www.ast.obs-mip.fr/grb/pz}}.\\
The accuracy of this redshift indicator and its use in this study is discussed 
in the following paragraph.  
\subsection{Justification of the pseudo-redshifts used in this study}
\label{justif}
\begin{figure}[t]
\begin{center}
\includegraphics[width=1\linewidth,angle=0]{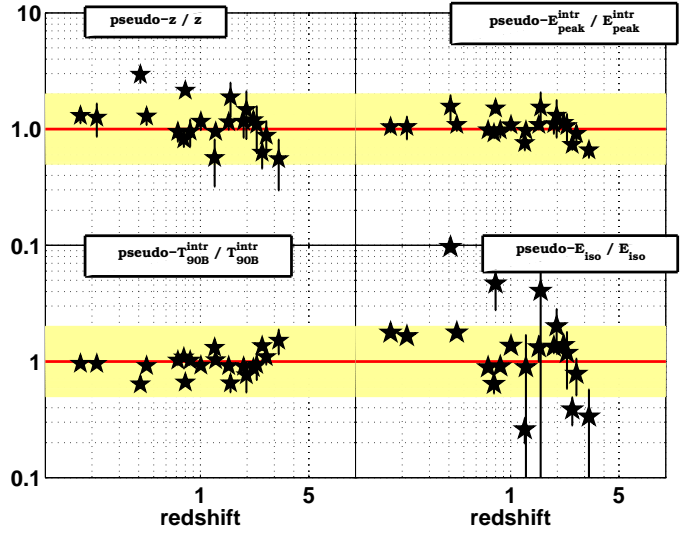}
\end{center}
\caption{\label{calib_allparams} Ratio between the {\it pseudo-redshift} 
and the redshift for the 20 GRBs contained in our sample 
for which both quantities are available (upper-left panel). 
The ratio of the 3 {\em pseudo-intrinsic} and {\em intrinsic} parameters 
studied are also shown, respectively \Ep\ (upper-right panel), $T_{90}$ 
(lower-left panel) and \Eiso\ (lower-right panel). 
In all plots, the filled region corresponds to a deviation from the equality 
between the two parameters by a factor of 2, 
and the error bars are at the 1$\sigma$ level (stat.+syst.)}
\end{figure}

As shown in Table~\ref{tab2}, 18 GRBs contained in our sample have 
a spectroscopic redshift and 2 have a photometric redshift (GRB~020127 
and GRB~030115). 
We computed a pseudo-redshift for the remaining 62 (among the 20 bursts 
having a known redshift, 14 were used for calibrating the pseudo-z).\\
In order to test the impact of the use of pseudo-redshifts in this work: 
\begin{itemize}
\item we computed the pseudo-redshifts of the 20 GRBs having 
a spectroscopic or photometric redshift,
\item for these same 20 GRBs we computed the ratio between the pseudo-redshift
and the redshift, and the ratio of the {\em pseudo-intrinsic properties}
derived with pseudo-redshifts to the {\em intrinsic properties} derived
using the redshift (Fig.~\ref{calib_allparams}).
\end{itemize}
We note that the dispersion of the ratio between the pseudo-redshifts
and the redshifts is smaller than a factor of 2, except for two outliers: GRB~020819
and GRB~051022 for which the spectroscopic redshift differs from the pseudo-redshift 
by factors of 2.95 and 2.15 respectively. For these 20 GRBs, 
the dispersion between the luminosity distance 
estimated with the pseudo-redshifts ($D_{L,estim}$) and 
the luminosity distance measured with the redshifts ($D_{L,meas}$) is: 
$\sigma_{D_{L}}=$~log$(D_{L,estim}/D_{L,meas})=$~0.125~dex, 
i.e a factor of 1.34\footnote{If we do not consider the two outliers GRB~020819 and GRB~051022 
the dispersion is $\sigma_{D_{L}}=$~0.089~dex, i.e a factor of 1.23.}.\\
The dispersion is also smaller than a factor of 2 for the \Ep\ 
(Fig.~\ref{calib_allparams}, top-right panel) 
and the $T_{90}$ parameters (Fig.~\ref{calib_allparams}, bottom-left panel). 
This is not surprising since these two quantities vary as $(1+z)$, reducing 
the impact of the redshift errors. 
\Eiso, which is more redshift-dependent than \Ep\ and the duration, is more scattered
and its dispersion is larger than a factor of 2. This corresponds to the bursts 
having the highest difference between their spectroscopic redshift 
and their pseudo-redshift: GRB~020819, GRB~040912B and GRB~051022 
have a pseudo-z estimate two to three times higher than their redshift; 
and GRB~020124, GRB~030115 and GRB~050408 
for which the pseudo-z is lower than the spectroscopic redshift by a factor of 2 
(Fig.~\ref{calib_allparams}, bottom-right panel).\\
We consider that the good agreement of pseudo-z with spectroscopic redshifts 
for the 20 GRBs having a measured $z$ is encouraging. One may question however 
if the pseudo-z values computed without knowing the spectroscopic redshift are as accurate as 
those computed when the redshift is known -- i.e. do pseudo-z have predictive 
power? -- Unfortunately, most of the bursts for which we have computed a pseudo-z 
do not have redshift measurements, so comparison with the redshift-estimates 
cannot be done. However, in some cases the pseudo-z values we issued in {\em GCN Circulars} 
were followed by a spectroscopic measurement. For instance, we determined the pseudo-redshifts 
of GRB~050525\footnote{The first {\em GCN Circular} containing the computation 
of a pseudo-redshift {\em in nearly real time} was done for this burst, using the 
spectral parameters obtained by Konus-WIND (Golenetskii et al \cite{golenetskii05a}). 
We found $\hat{z}=0.36\pm0.10$ (Atteia \& P\'elangeon \cite{atteia05b}) 
whereas the spectroscopic redshift obtained by spectroscopy of the host galaxy 
and published the day after was $z=0.606$ (Foley et al. \cite{foley05}). 
The difference between the pseudo-redshift and the redshift 
was due to an incorrect published fluence (Golenetskii et al. \cite{golenetskii05b}). 
As soon as the correct value was available, we recomputed the pseudo-redshift 
of this burst and found $\hat{z}=0.64\pm0.10$ (Atteia \& P\'elangeon \cite{atteia05c}).} 
and GRB~070125\footnote{We have computed a redshift-estimate of this bright burst 
using the spectral parameters derived from two different instruments, 
Rhessi ($\hat{z}=1.63\pm0.80$, P\'elangeon \& Atteia \cite{pelangeon07a}) 
and Konus-WIND ($\hat{z}=1.34\pm0.30$, P\'elangeon \& Atteia \cite{pelangeon07b}). 
Both values were consistent with a redshift of $\sim$1.5. 
The spectroscopic redshift $z\sim1.54$ was measured several days 
after our estimate (Fox et al. \cite{fox07}; see also Cenko et al. \cite{cenko08}).} 
before their redshifts were known. The spectroscopic measurements have confirmed 
our estimates. Moreover, two other GRBs detected in 2003 and 
2004 by HETE-2 had a pseudo-redshift (P\'elangeon et al. \cite{pelangeon06a}) 
but no redshift. Some later observations of the host galaxies 
resulted in the determination of their spectroscopic redshifts, 
in agreement with their pseudo-redshifts: Rau et al. (\cite{rau05}) measured the redshift 
of the host galaxy of GRB~030531 to be $z=0.782\pm0.001$ 
($\hat{z}=0.64\pm0.15$, P\'elangeon et al. \cite{pelangeon06a}),  
and Stratta et al. (\cite{stratta07}) obtained the redshift of the X-ray flash 040912: 
$z=1.563\pm0.001$ from the [OII] line of the host-galaxy spectrum, 
also consistent with its previously determined pseudo-redshift 
($\hat{z}=2.90\pm1.60$, P\'elangeon et al. \cite{pelangeon06a}).\\
\\
In order to quantify the effect of the uncertainty from the pseudo-redshifts 
on the intrinsic distributions, we performed the following test: 
we artificially increased the errors on the pseudo-z values 
which are taken into account in the construction of the intrinsic 
parameters distributions (see Section~\ref{Epi}). 
This extra dispersion takes on three values corresponding 
to factors of 1.4, 2.0 and 3.0. 
We tested the impact of this extra dispersion 
on the intrinsic \Ep\ distributions (`simple' and  `unbiased'), 
and thus we show that even if the pseudo-z were only accurate 
to a factor of 2 or 3, this would not significantly change the intrinsic parameter 
distributions and all the subsequent results obtained in this paper (see Sect.~\ref{s4}).\\
\\
It was recently proposed by Butler et al. (\cite{butler07}) 
that the correlations between the intrinsic GRB parameters found 
in the pre-Swift era were due to instrumental biases 
(e.g. detector thresholds effects) rather than to real physical properties 
of the sources. This is the result of a study they performed 
on a sample of 77 Swift GRBs with measured redshifts, where they found that 
those GRBs are on average harder than the pre-Swift measurements, i.e. they 
have higher \Ep\ and lower \Eiso. As a consequence, they conclude 
that these correlations cannot be used to estimate the GRB redshifts.\\ 
Since no clear conclusion can be drawn from this debate, we have also considered 
this possibility and we calculated all the distributions twice: 
one case where we attribute a pseudo-redshift to the 62 bursts which 
have no redshift, and the other case where we randomly attribute one of the 24 
secure redshifts measured for the HETE-2 long GRBs. Therefore, for all the distributions 
we obtain we will comment on how they change between the two cases.\\ 
\\
Throughout the paper we use the following notations: 
\begin{itemize}
\item the spectroscopic or photometric redshifts are denoted by $z$. 
\item the pseudo-redshifts are denoted by $\hat{z}$. 
\item $z_{sample}$ refers to the value of the redshift used in our study: 
this is the spectroscopic/photometric redshift ($z$) when it is available 
and the pseudo-redshift ($\hat{z}$) otherwise (see Table~\ref{tab1}). 
\end{itemize}
%
%
%
\section{Parameter 1: \Ep}
\label{s4}
\subsection{Measuring the \Ep\ of X-ray flashes}
The peak energy of a burst is generally measured using the {\em XSPEC} software 
(Arnaud \& Dorman \cite{arnaud03}) and fitting a phenomenological model 
-- {\em GRB (Band) Model} (Band et al. \cite{band93}) or {\em Cutoff Power Law Model} -- 
on the spectral data. However, for some X-ray flashes, the peak energy 
is lower than 10~keV, so the data do not constrain the four parameters of the 
Band function. For those bursts, the value obtained is too close to the lower boundary 
of FREGATE (6~keV) and of the WXM (2~keV). In this case, we observe a power-law spectrum 
instead of a Band function. To solve this problem, Sakamoto et al. (\cite{sakamoto04}) 
showed that a modified function, called the {\em Constrained Band Model (CBM)}, 
based on a three-dimensional subspace of the full four-dimensional Band function 
parameter space, could be used. We adopted this model to fit the data 
when it was necessary (see Table~\ref{tab1}).\\
However, in order to check whether such a model could add a bias 
to the estimate of \Ep, we used XSPEC to simulate an XRF spectrum as follows: 
\begin{itemize}
\item we use the data of a real XRF, in order to have real background 
statistics and instrumental response matrices. 
\item we construct a fake burst that follows a Band model with typical values 
of -1.0 and -2.3 for the power-law indices at low and high energy respectively. 
These parameters correspond to the mean values for GRB spectra 
according to BATSE results (see e.g. Preece et al. \cite{preece00}; 
Kaneko et al. \cite{kaneko06}). The two other parameters, \Ep\ and the normalization, 
are the variable parameters that let us produce fake XRFs of greater or lesser intensity. 
\item we generate fake data files for XRFs of various brightnesses 
and various \Ep.
\item we then perform a spectral analysis using a Band model and a Constrained Band model, 
and compare the parameters obtained. 
\end{itemize}
We find that the Band model is valid for \Ep\ down to 8~keV. For this limiting value, 
the errors on the spectral parameters are slightly larger than those 
obtained using the Constrained Band function, but they are still valid. For \Ep\ below 8~keV, 
the Constrained Band model is preferable: it gives correct \Ep\ values 
or provides at least reliable upper limits.\\
In all cases we obtain consistent values between the CBM and the Band function 
when both functions are applicable, and these values are consistent 
with the ones we introduced to produce the fake XRFs. We thus conclude that 
the values determined with the Constrained Band model are not biased.
\subsection{The observed \Ep\ distribution}
\label{Epo}
\begin{figure}[ht]
\begin{center}
\includegraphics[width=1\linewidth,angle=0]{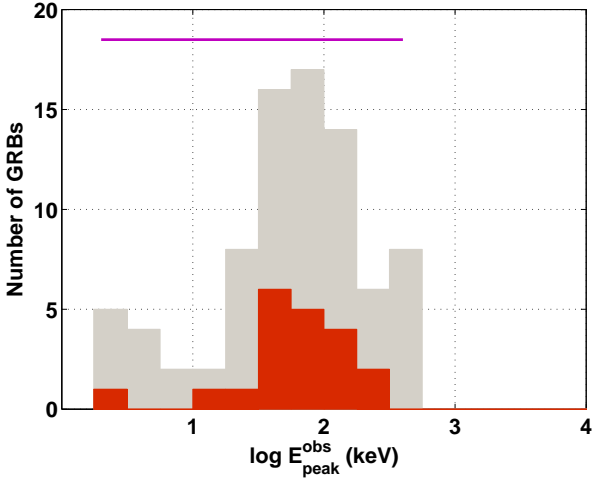}
\end{center}
\caption{\label{dist_Epo} Distribution of the \Epobs\ measured 
for the complete sample of 82 GRBs detected and localized by HETE-2. 
The sub-sample's histogram of bursts with secure redshifts ($z$) is shown in red, 
and the total sample's histogram is shown in grey. The bursts on the left part of 
the histogram correspond to the events that have a measured higher limit in \Epobs\ 
of about 4~keV (see Table~\ref{tab1}). The energy range of HETE-2 (2-400~keV) 
is symbolized by the magenta horizontal solid line.}
\end{figure}

Several points can be noticed in Figure~\ref{dist_Epo}: first, the \Epobs\ distribution peaks at about 100~keV 
and extends over the entire energy range of HETE-2 (2-400~keV). 
Second, the two classes of bursts (classical GRBs and X-ray flashes) are clearly distinguished. 
As a comparison, the study of Kaneko et al. (\cite{kaneko06}) 
on the complete spectral catalog of bright BATSE Gamma-Ray Bursts, leads to 
an \Epobs\ distribution peaking at about 300~keV, extended at high energies up to several MeV 
and without significant events at very low energies. The paucity of soft ($\sim$a few keV) 
and hard ($\sim$MeV) bursts in our sample is probably not real. For the soft bursts, 
this is clearly visible by a simple comparison of the BATSE and HETE-2 \Ep\ distributions. 
As BATSE triggers were generally done on the count rate between 50 and 300~keV, 
the ratio between the XRFs and the GRBs is low. For HETE-2, the energy range 
extends down to 2~keV and enables the detection of more XRFs but still 
prevents the detection of XRFs with \Ep\ below a few keV. As far as the hard tail 
of the \Ep\ distribution is concerned, the low sensitivity 
of the detectors at high energy is also one of the causes 
of the inefficient detection of hard bursts. Moreover, two bursts emitting 
the same energy with different hardness do not have the same number of photons: 
the softer GRB has more photons than the harder one. 
Consequently, BATSE-like or HETE2-like missions are probably not able to derive 
a true observed \Ep\ distribution. The distributions obtained clearly show 
that the \Epobs\ distribution depends strongly on the energy range 
of the instruments involved.
\subsection{The intrinsic \Ep\ distribution}
\label{Epi}
\begin{figure}[ht]
\begin{center}
\includegraphics[width=1\linewidth,angle=0]{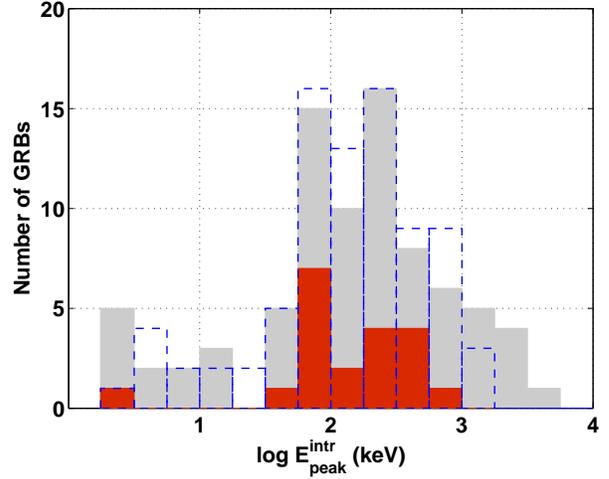}
\end{center}
\caption{\label{dist_Epi} Distribution of \Epi\ for the complete sample of 82 GRBs 
detected and localized by HETE-2. The histogram related to the sub-sample of bursts with secure redshifts ($z$) 
is shown in red, and the total sample's histogram is shown in grey. The blue dashed 
histogram corresponds to the studied case 2 (see Part~\ref{justif}), i.e. 
the sample of 62 GRBs without secure redshift that are randomly given a redshift 
from the HETE-2 redshift distribution.}
\end{figure}

So far, the intrinsic \Ep\ distribution has been very little discussed, 
mainly due to the difficulty of measuring both the \Ep\ and the redshift 
of the bursts. On the one hand, BATSE, Beppo-SAX and HETE-2 did constrain 
the \Epobs\ thanks to their wide energy ranges, but the bursts they 
detected suffered from a lack of redshift determination. On the other hand, 
thanks to its performance and fast localization capability, Swift currently 
enables more redshift measurements, but the narrow energy range 
of the BAT instrument (15-150~keV) does not allow a good constraint 
of the \Epobs.\\
As mentioned by Amati (\cite{amati06}), about 70 intrinsic \Ep\ are available, 
but this sample contains bursts detected by different satellites. 
Our work is based on a sample containing all the bursts localized with a single 
satellite, hence avoiding selection effects inherent to the use of 
data from different instruments.\\ 
We see in Figure~\ref{dist_Epi} that the \Epi\ distribution 
is broader than the \Epobs\ distribution. 
At high energies, some bursts have an \Epi\ reaching a few MeV. Nevertheless, 
as we previously mentioned, the energy range of HETE-2 (2-400~keV) 
prevents us from drawing any conclusion about the high energy part 
of the \Ep\ distribution.\\
At low energies, bursts having an \Epobs\ lower than 20~keV, 
and defined as X-ray flashes in the observer frame, 
still have a low \Ep\ in their source frame. 
For most of those GRBs, the low intrinsic peak energy 
is due to their intrinsic faintness 
as they are at low or intermediate redshift. 
(This point is discussed in more detail in Section~\ref{s8}).\\
Joining these two extreme energy ranges, the intrinsic \Ep\ distribution 
of HETE-2 GRBs extends over 3 decades in energy. 
The true width of the distribution is probably even larger because bursts 
with \Ep\ lower than 1~keV are not detected, 
and bursts with \Ep\ greater than 1~MeV cannot be constrained 
by an HETE2-like mission. 
\subsection{The unbiased \Ep\ distribution}
\label{Epu}
\subsubsection{Construction}
In order to derive the {\em true \Ep\ distribution}, the distance-scale is not 
the only correction that has to be applied to the observed \Ep\ distribution. 
All the bursts do not have the same brightness and are not detected 
at the same significance in the observer frame. We took this into account to correct 
the intrinsic \Ep\ distribution obtained in the previous section.\\ 
To do this we used the following method:\\ 
\\
(1) in order to take into account the errors on all the parameters used 
(spectral properties and redshift) we have computed `smoothed distributions' by:
\begin{itemize}
\item producing for each burst a set of 100 $E_{peak,sim}^{obs}$ 
randomly selected within the 90\% confidence level (c.l.) error range 
of the measured \Epobs\, given in Table~\ref{tab1}\footnote{For some XRFs, 
the lower error limit could not be constrained. In this case the value was set 
to $E_{peak,min}^{obs}=$~1~keV. For some other GRBs, 
it was the upper error limit which was not constrained. 
In this case we set the value to $E_{peak,max}^{obs}=$~1~MeV.} and 100 redshifts 
randomly selected within the 90\%~c.l. error range 
of their corresponding spectroscopic/photometric redshift 
or pseudo-redshift $z_{sim}$. Combined with the 100 simulated 
$E_{peak,sim}^{obs}$, we thus obtained a sample 
of 100 simulated $E_{peak,sim}^{intr}$.
\item representing each of the 8200 simulated bursts ($b$) 
with a normal distribution of the logarithmic \Epi\ values ($LE_{p,b}^{intr}$): 
\begin{equation}
f_{b}(le_{p,i})=\frac{1}{\sigma_{LE_{p,b}^{intr}}}\times \exp\biggl[-\frac{1}{2}~\biggl(\frac{le_{p,i}-\mu_{LE_{p,b}^{intr}}}{\sigma_{LE_{p,b}^{intr}}}\biggr)^2\biggr]
\end{equation}
where $le_{p,i}$ is the variable of the functions $f_b$, $\mu_{LE_{p,b}^{intr}}$ 
is the base 10 logarithm of $E_{peak,sim}^{intr}$ related to the burst $b$ 
and $\sigma_{LE_{p,b}^{intr}}$ is set to 0.05. 
\end{itemize}
(2) then, we computed for the 8200 simulated GRBs 
a weight related to their detection significance. For that, we:
\begin{itemize}
\item determine the maximum redshift $z_{max}$ at which the source could have been 
detected by the instruments, by first comparing the SNR computed in Section~\ref{s2} 
(see Table~\ref{tab1}) to the SNR threshold of the HETE-2 instruments, and 
then by redshifting the sources until their peak photon flux reaches the SNR threshold 
of the trigger instrument, giving us both the visibility distance between 
the source and the satellite and the maximum redshift $z_{max}$.
\item assume that the GRB rate follows the star formation rate. 
For this we have adopted the model SFR$_2$ of Porciani \& Madau (\cite{porciani01}) 
that reproduces a fast evolution between $z=0$ and $z=1$ 
and remains constant beyond $z\ge2$
\begin{equation}
R_\mathrm{SFR_2}(z)\propto\,0.15 h_{65} \frac{exp(3.4z)}{exp(3.4z)+22}
\end{equation}
\item derive for each burst the number of GRB per year within its visibility volume
\begin{equation}
\label{NV}
N_{Vmax}\propto\,\int_0^{z_{max}} dz\,\frac{dV(z)}{dz}\,\frac{R_{SFR_2}(z)}{1+z}
\end{equation} 
In this equation $dV(z)/dz$ is the comoving element volume, described by 
\begin{equation}
\frac{dV(z)}{dz}=\frac{c}{H_0}\,\frac{4\pi\,dl^2(z)}{(1+z)^2\,[\Omega_{M}(1+z)^3+\Omega_{K}(1+z)^2+\Omega_{\Lambda}]^{1/2}}
\end{equation}
where $H_0$ is the Hubble constant, $\Omega_{K}$ is the curvature contribution 
to the present density parameter ($\Omega_{K}=$~1$-\Omega_{M}-\Omega_{\Lambda}$), 
$\Omega_{M}$ is the matter density and $\Omega_{\Lambda}$ is the vacuum density. 
Throughout this paper we have assumed a flat $\mathrm{\Lambda}$CDM universe where 
($H_0$, $\Omega_{M}$, $\Omega_{\Lambda}$)$=$(65$h_{65}$~km~s$^{-1}$Mpc$^{-1}$, 0.3, 0.7).\\ 
This procedure allows us to give each burst a weight ($W_b$) inversely proportional to $N_{Vmax}$. 
The rationale of weighting each burst by $1/N_{Vmax}$ is the following: 
the visibility volume is different for each GRB of our sample. Moreover, 
each burst observed is randomly taken from all the bursters present 
in its visibility volume. In this way, rare bright bursts, having a large 
visibility volume, will have low weights, while faint local GRBs 
will have heigher weights. This procedure also takes into account the fact 
that the GRB rate evolves with redshift, leading to the fact that GRBs 
are about ten times more frequent at $z\sim1$ than at present. Note that 
an evolution of the GRB distributions with the redshift is not included in 
this procedure, although such evolution probably exists (e.g. Daigne, Rossi 
\& Mochkovitch \cite{daigne06}). This case is not addressed here because we do not have 
a sufficient number of GRBs.
\end{itemize}
(3) the total smoothed distribution (SD) is the sum 
of the individual functions $f_{b}(le_{p,i})$, 
normalized by the corresponding burst's weight $W_b$
\begin{equation}
\rm{SD}_{(LE_{peak,sample}^{intr})}=\frac{1}{n_{simu}}\times \sum_{b=1}^{n_{burst}}W_b\times\,f_b(le_{p,i})
\end{equation}
with $n_{simu}$ the number of simulations (100) for each burst 
and $n_{burst}$ the number of bursts contained in our sample (82).
\subsubsection{Results}
\begin{figure}[t]
\begin{center}
\includegraphics[width=1\linewidth,angle=0]{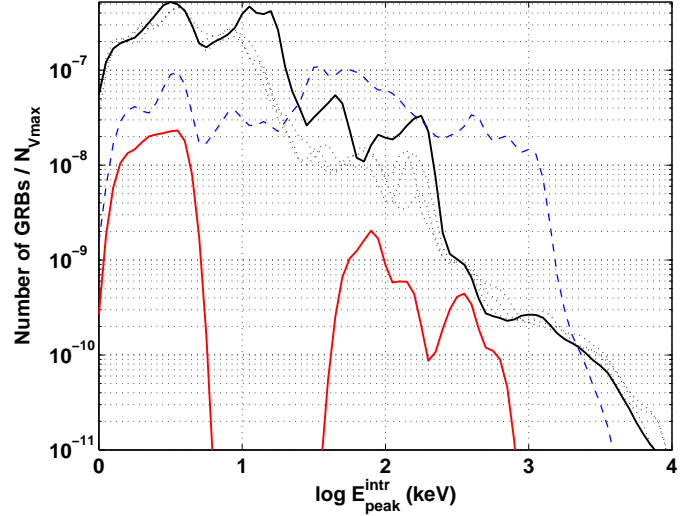}
\end{center}
\caption{\label{dist_Epu} Unbiased distributions of the intrinsic \Ep\ 
for the complete sample of 82 GRBs detected and localized by HETE-2. 
The smoothed distribution (SD) based on the 20 bursts having 
a spectroscopic/photometric redshift ($z$) 
is represented by the lower solid red curve and the one corresponding 
to the complete sample is shown with the upper solid black curve. 
The three black dotted smoothed distributions show the test performed 
to quantify the impact of the pseudo-redshift accuracy in this study 
(all the pseudo-redshifts are given an error range 
artificially increased by factors of 1.4, 2.0 and 3.0). 
The blue dashed SD corresponds to case~2 (see Part~\ref{justif}), 
i.e. the sample of 6200 simulated GRBs (corresponding to the 62 GRBs 
of our sample without secure redshift) which are randomly given a redshift 
from the HETE-2 redshift distribution.}
\end{figure}

The resulting unbiased \Ep\ distribution for our sample of bursts 
with this method is shown in Figure~\ref{dist_Epu}. 
In this Figure we see that the \Ep\ distribution has dramatically changed, 
with a clear domination of the bursts having low intrinsic \Ep, 
i.e. the X-ray flashes.\\ 
As explained in Section~\ref{s3}, we tested whether a high accuracy 
for the pseudo-redshifts was necessary, i.e. whether the results obtained 
could depend on these estimated distance scales. To do this, we performed 
the same analysis as described throughout this Section, but this time 
for simulated redshifts randomly selected within the error ranges based on the 
initial ones and artificially increased by factors of 1.4, 2.0 and 3.0. 
We find that both the distribution of intrinsic \Ep\ 
and the unbiased distribution of intrinsic \Ep\ do not significantly depend 
on the accuracy of the pseudo-z within a factor~2 or 3 (Fig.~\ref{dist_Epu}). 
This result can be explained by the intrinsic dispersion of the parameters 
studied which is so large that the uncertainties in the distance scales 
do not strongly affect the results.\\
Moreover, we also performed this analysis by assuming for the distance 
of each of the 62 GRBs without secure redshift a set of 100 redshifts 
randomly taken from the total redshift distribution 
of HETE-2 long bursts (24 values). The results obtained in this case 
differ significantly because the predominence of low \Epi\ bursts 
has disappeared. With this method, the distribution is flat 
and broad. Here again, the distribution's width may not reflect 
the true one, but it contributes interesting information 
to the unbiased \Ep\ distribution, which is probably situated 
-- `in reality' -- between the two curves presented in Fig.~\ref{dist_Epu} 
(black solid line and blue dashed line).
%
%
%
\section{Parameter 2: $T_{90}$}
\label{s5}
In this section we study the duration distribution 
of the complete sample of HETE-2 GRBs. We adopted $T_{90}$  
as the common definition for all the bursts. We derive in the following 
sub-sections the observed, intrinsic and unbiased log~$T_{90}$ 
distributions using the same method as described 
for the study of the intrinsic log~\Ep\ distribution (Sect.~\ref{s4}).
\subsection{The observed $T_{90}$ distribution}
\begin{figure}[ht]
\begin{center}
\includegraphics[width=1\linewidth,angle=0]{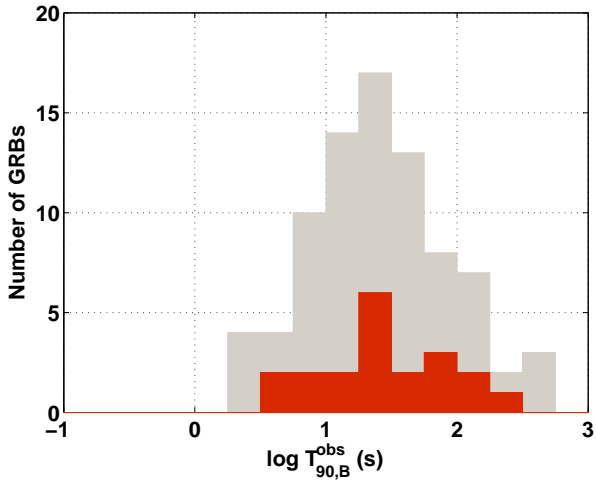}
\end{center}
\caption{\label{dist_tBo} Distributions of the log $T_{90}$ measured 
in the observer frame and in band $B$ of FREGATE (6-80~keV) 
for the complete sample of 82 GRBs localized by HETE-2. 
The histogram related to the sub-sample of bursts with secure redshifts ($z$) 
is shown in red, and that of the total sample is shown in grey.}
\end{figure}

$T_{90}$ is defined as the duration for which 90\% of the counts 
in a given energy range are detected. Recall that in this study 
we only focus on long-duration GRBs. Over the 4 FREGATE bands 
($A$: 6-40~keV; $B$: 6-80~keV; $C$: 30-400~keV; $D$: $>$400~keV) 
we chose the one that contained both the classical GRBs and the XRFs, 
i.e. the band $B$ (6-80~keV). The log~$T_{90}$ distribution 
in this energy range (see Fig.~\ref{dist_tBo}) extends from 0.33 to 2.73. 
The mean value of this distribution is 1.40 and the median value 1.33.
\subsection{The intrinsic $T_{90}$ distribution}
\begin{figure}[ht]
\begin{center}
\includegraphics[width=1\linewidth,angle=0]{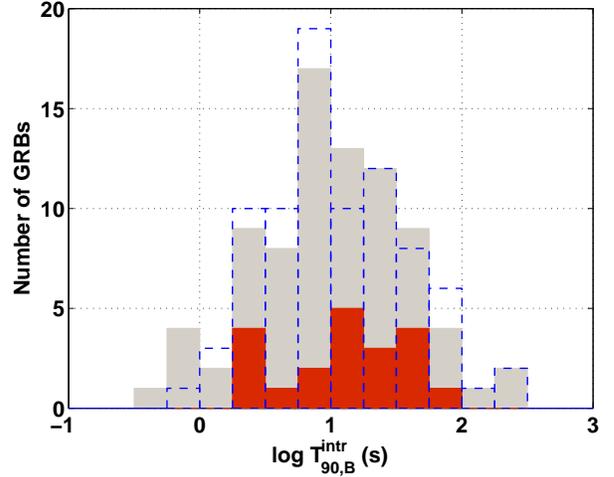}
\end{center}
\caption{\label{dist_tBi} Intrinsic log~$T_{90}$ distributions 
for the complete sample of 82 GRBs detected and localized by HETE-2. 
The histogram of the sub-sample of bursts with secure redshifts ($z$) 
is shown in red, and the total sample histogram is shown in grey. 
The blue dashed histogram corresponds to the case~2 (see Part~\ref{justif}), 
i.e. to the 62 GRBs without secure redshift that were randomly given a redshift 
from the HETE-2 redshift distribution.}
\end{figure}

The histograms are shown in Figure~\ref{dist_tBi}.
We caution that the observed $T_{90}$ are measured in the same energy range 
in the observer frame (see Table~\ref{tab1}). Since this energy range depends 
on the redshift of the source, it is different for each burst and this has 
an impact on the intrinsic duration measurement.\\
The histogram of the intrinsic log~$T_{90}$ in the (6-80~keV) energy range extends 
from -0.26 to 2.40. The mean value of this distribution is 1.01 and the median value is 1.00.\\
Contrary to the log~$E_\mathrm{peak}$ distribution, the intrinsic log~$T_{90}$ distribution 
is not broader than the observed log~$T_{90}$ distribution and is extended over about 
the same range ($\sim$2.5~decades).
\subsection{The unbiased $T_{90}$ distribution}
\label{T90ia}
\begin{figure}[ht]
\begin{center}
\includegraphics[width=1\linewidth,angle=0]{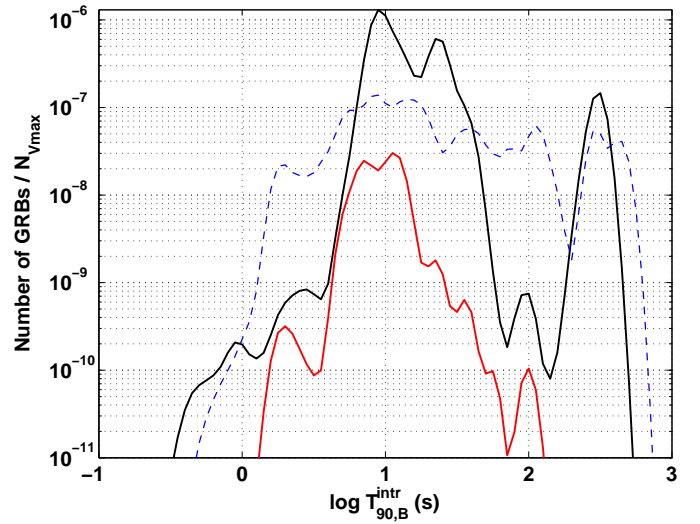}
\end{center}
\caption{\label{dist_tBu}  Unbiased distributions of the intrinsic log~$T_{90}$ 
for the complete sample of 82 GRBs detected and localized by HETE-2. 
The smoothed distribution based on the 20 bursts with spectroscopic/photometric 
redshifts ($z$) 
is represented by the lower solid red curve; the one for the complete sample 
is given by the upper solid black line. 
The blue dashed SD corresponds to case~2 (see Part~\ref{justif}), 
i.e. the sample of 6200 simulated GRBs 
(corresponding to the 62 GRBs of our sample without secure redshift) 
which were randomly given a redshift from the HETE-2 redshift distribution.}
\end{figure}

Correcting the intrinsic $T_{90}$ for the $N_{Vmax}$ (see Part~\ref{Epu}) 
we obtained the unbiased $T_{90}$ distribution (Fig.~\ref{dist_tBu}).\\
We note that the GRB population is dominated by events with an intrinsic duration 
of about 10~s (for an energy range of 6-80~keV in the observer frame). 
Nevertheless, a non-negligable number of GRBs 
have a long intrinsic duration lasting about 300~s.\\
If we consider the case~2 study (Fig.~\ref{dist_tBu}), 
we obtain a significantly different unbiased $T_{90}$ distribution, 
more flat than the one obtained in case~1, 
i.e. no typical intrinsic duration for the GRB appears 
when considering distances randomly taken within 
the total redshift distribution of HETE-2 GRBs. 
%
%
%
\section{Parameter 3: isotropic energy}
\label{s6}
\subsection{The observed fluence}
\begin{figure}[ht]
\begin{center}
\includegraphics[width=1\linewidth,angle=0]{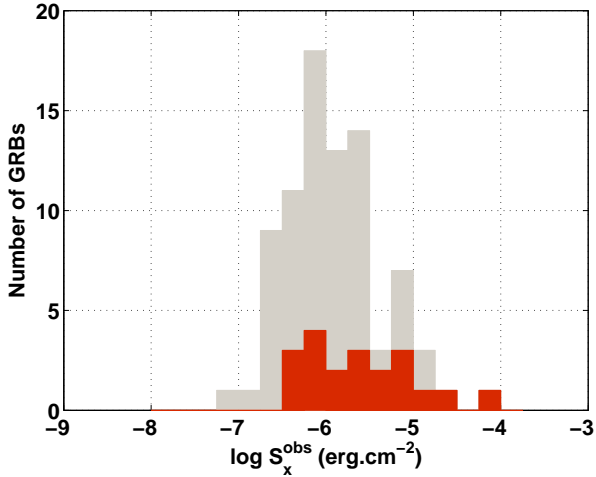}
\end{center}
\caption{\label{dist_sx} Distributions of the logarithmic fluence 
in the $X$ energy range (2-30~keV) for the complete sample of 82 GRBs localized by HETE-2.
The histogram for the sub-sample of bursts with secure redshifts ($z$) 
is shown in red, and the total sample histogram is shown in grey.}
\end{figure}

Taking the fluences (i.e. the fluxes integrated over the duration of the bursts' spectra) 
in the HETE-2 Catalog (Vanderspek et al. \cite{vanderspek08}) and applying 
the same method described in Section~\ref{s4}, we show in Figure~\ref{dist_sx} 
the histograms of the logarithmic fluence in the energy range 2-30~keV. 
The distribution of the log~$S_X$ extends between -7.12 and -4.24. 
The mean and median values of this distribution are -5.90 and -6.00, respectively.
\subsection{The isotropic energy}
\begin{figure}[ht]
\begin{center}
\includegraphics[width=1\linewidth,angle=0]{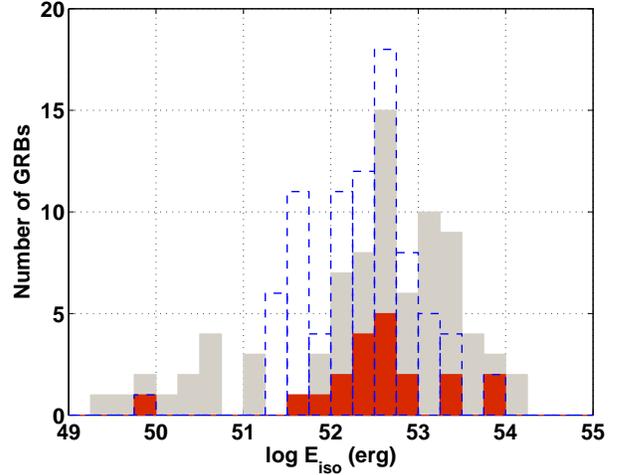}
\end{center}
\caption{\label{dist_eiso} \Eiso\ distributions for the complete 
sample of 82 GRBs localized by HETE-2.
The histogram of the sub-sample of bursts with secure redshifts ($z$) 
is shown in red, and the histogram of the total sample 
is shown in grey. The blue dashed histogram corresponds to case~2 
(see Part~\ref{justif}), i.e. to the sample of 62 GRBs without secure redshifts 
that were randomly given a redshift from the HETE-2 redshift distribution.}
\end{figure}

The equivalent of the fluence measured in the observer frame is the 
energy radiated in the source frame. One easy way 
to estimate this energy is to assume an isotropic emission (\Eiso) integrated 
over a fixed energy-range (1-10$^4$~keV in the observer frame, 
see e.g. Amati et al. \cite{amati02}, Amati \cite{amati06}). 
We computed this quantity for the 82 bursts and obtained 
the distribution shown in Fig.~\ref{dist_eiso}. This distribution 
of log~\Eiso\ extends from 49.50 to 54.24, and has a mean value of 52.44 
and a median value of 52.62. Two classes of bursts are evident, 
with a separation at about 51.5. Most of the bursts have a log~\Eiso\ 
lying between 52 and 54 (the classical GRBs) and the second sample 
has a log~\Eiso\ lower than 51.
\subsection{The unbiased isotropic energy}
\label{lumia}
\begin{figure}[ht]
\begin{center}
\includegraphics[width=1\linewidth,angle=0]{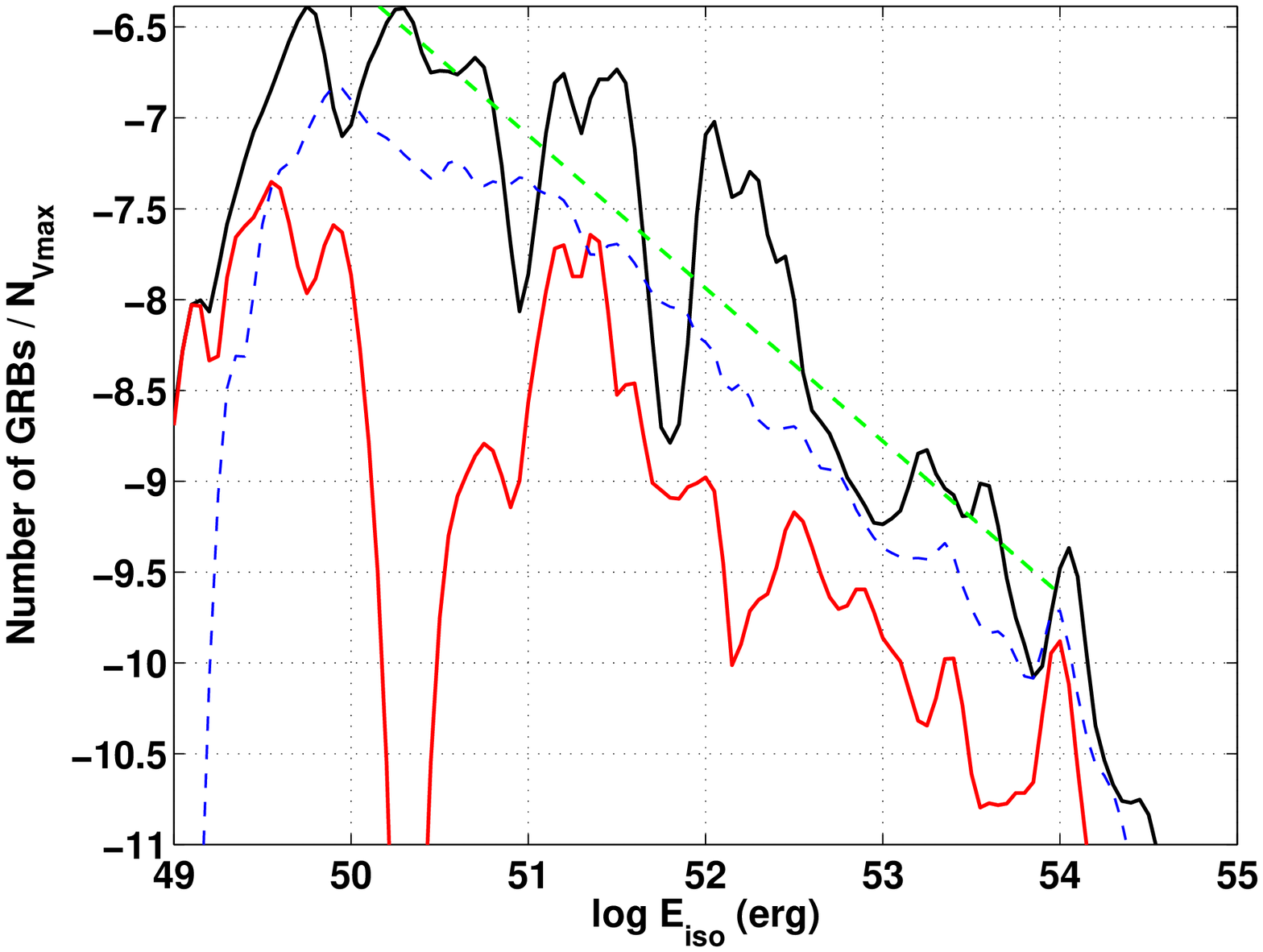}
\end{center}
\caption{\label{dist_eisou} Unbiased \Eiso\ distributions for the complete sample 
of 82 GRBs localized by HETE-2. 
The smoothed distribution based on the 20 bursts with a spectroscopic/photometric 
redshift ($z$) is represented by the lower solid red curve; 
the one for the complete sample is shown by the upper solid black line. 
The blue dashed SD corresponds to case~2 (see Part~\ref{justif}), 
i.e. the sample of 6200 simulated GRBs (corresponding to the 62 GRBs of our sample 
without secure redshifts) which were randomly given a redshift from the HETE-2 
redshift distribution. The best-fit power-law (green dashed line) for the complete sample, 
estimated between $E_{iso}=$10$^{50}$ and 10$^{54}$~erg is 
$10^{-7.94}\times\big( \frac{E_{iso}}{10^{52}}\big)^{-0.84}$.}
\end{figure}

The correction of these \Eiso\ for the $N_{Vmax}$ of each burst gives us 
the {\em unbiased} \Eiso\ distribution (see Fig.~\ref{dist_eisou}). 
This distribution decreases with \Eiso, and the best-fit power-law 
for the complete sample distribution, estimated between \Eiso$=$10$^{50}$~erg 
and \Eiso$=$10$^{54}$~erg is 10$^{-7.94}\times\,E_{iso,52}^{-0.84}$, 
where $E_{iso,52}$ is in units of 10$^{52}$~erg.
As for the two previous cases studied, this distribution is dramatically 
different from the observed and intrinsic ones. It shows a clear predominence 
of bursts with low \Eiso. This distribution hence strengthens 
the result we obtained in Section~\ref{s4}, i.e. the predominence 
of X-ray flashes in the overall GRB population. Here again we can discuss 
the impact of our second case calculation. In this case, the shape 
of the distribution is similar to the one obtained in case~1.
%
%
%
\section{Correlations}
\label{s7}
\begin{figure*}[!htb]
\begin{center}
\includegraphics[width=1\linewidth,angle=0]{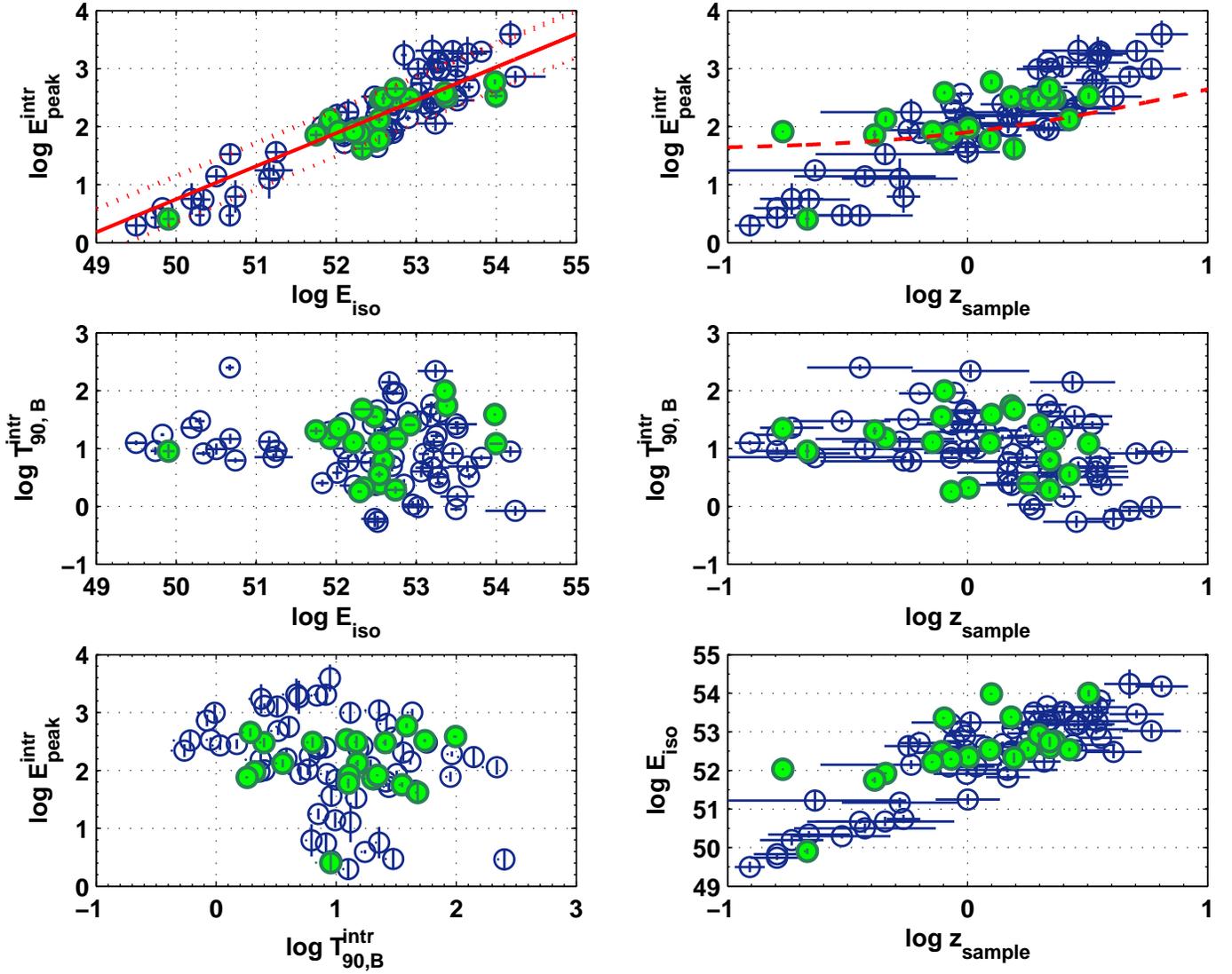}
\end{center}
\caption{\label{correl} Correlations between the intrinsic parameters studied: 
\Epi, $T_{90}^{intr}$, \Eiso\ and the distance-scale \zsamp. 
In each panel the GRBs that have a measured redshift ($z$) are represented with the green 
filled circles, and the ones having a pseudo-redshift ($\hat{z}$) are symbolized with the 
blue open circles. In the top left panel, by way of comparison, we show (solid line) 
the best-fit power-law \Epi$=$77$\times\,_{iso,52}^{0.57}$ 
(with \Epi\ in keV and $E_{iso,52}$ in units of 10$^{52}$~erg) found by Amati (\cite{amati06}) 
and also the vertical logarithmic deviation of 0.4 (dotted lines) displayed in Amati (\cite{amati06}). 
In the top right panel the dashed curve marks the redshift evolution of the \Ep, 
assuming a burst with $E_{peak}^{(z=0)}=$~40~keV (see Stratta et al. \cite{stratta07}).}
\end{figure*}

Having determined the intrinsic properties of the HETE-2 GRBs, 
we test whether any correlations are found between these values. 
We performed a pairwise comparison of the four intrinsic parameters 
(\Ep, $T_{90}$, \Eiso, \zsamp) to search for possible correlations. 
Six plots are obtained and shown in Figure~\ref{correl}. The corresponding correlation 
coefficients and associated $p$-values testing the hypothesis of no 
correlation are presented in Table~\ref{corr_coeff}.
\begin{table}[h!]
\caption{\label{corr_coeff} Correlation coefficients ($R$) obtained for the 
pairwise comparison of the four intrinsic parameters discussed in this 
paper (\Ep, $T_{90}$, \Eiso, \zsamp). The lower and upper bounds are at a 95\% 
confidence interval. The corresponding $p$-values test the non-correlation 
by computing the probability of having a correlation as large as the 
observed value by chance.}
\begin{center}
\begin{tabular}{l l l}
\hline\hline
Tested correlation & $R$ & $p$-values\\
\hline
\Eiso-\Epi   & 0.912$_{-0.046}^{+0.031}$  & 1.215$\times$10$^{-32}$\\
\zsamp-\Epi  & 0.860$_{-0.070}^{+0.048}$  & 4.297$\times$10$^{-25}$\\
\Eiso-\T90i  & -0.102$_{-0.211}^{+0.220}$ & 3.596$\times$10$^{-1}$\\
\zsamp-\T90i & -0.354$_{-0.177}^{+0.206}$ & 1.087$\times$10$^{-3}$\\
\T90i-\Epi   & -0.255$_{-0.193}^{+0.215}$ & 2.060$\times$10$^{-2}$\\
\zsamp-\Eiso & 0.831$_{-0.083}^{+0.057}$  & 4.729$\times$10$^{-22}$\\
\hline
\end{tabular}
\end{center}
\end{table}
We note that the \Epi--\Eiso\ (Amati) correlation is found (1$^\mathrm{st}$ panel) 
and that the sub-sample of XRFs fills the gap between 
XRF~020903 and the cluster of intrinsic classical GRBs (see next Section). 
This is not really surprising since our pseudo-redshift estimates assume 
that the \Epi--\Eiso\ relation is valid. It is nevertheless interesting 
to note that nearly all of the 82 GRBs in our sample are consistent 
with this correlation with no strong outlier. While it is not the purpose 
of this paper to discuss the reality of \Epi--\zsamp\ and \Eiso--\zsamp\ correlations, 
we note that these correlations could be partially or totally explained by 
selection effects which practically prevent the detection of low-luminosity GRBs 
at high redshift. In the three last plots, no other 'simple' 
tight correlation is found, but these plots interestingly show the instrumental limits 
affecting the detection of the bursts (duration, energetics and redshift). 
For instance, we can clearly see that HETE-2 was lucky to detect the bright 
GRB~030329 (see panels~2 and 6). Hence Figure~\ref{correl} can be used to show 
the characteristics of the bursts missed by HETE-2, i.e. mainly the GRBs 
at low-redshift with high \Epi\ or high \Eiso; and conversely the GRBs 
with low \Ep\ occuring at high redshift. 
%
%
%
\section{The nature of X-ray flashes}
\label{s8}
The demonstration of the existence of X-ray flashes (Heise et al. \cite{heise01}; 
Kippen et al. \cite{kippen01}) was immediately followed by several theories attempting 
to explain their nature 
and to determine whether they belong to the same class as the classical GRBs, 
or whether they are different cosmological events. Thanks to various studies 
of samples of X-ray flashes, several clues now allow us to unambiguously 
associate XRFs with the classical GRBs. XRFs are GRBs with lower \Ep\ 
and higher fluxes in $X$-rays than in $\gamma$-rays (e.g. Kippen et al. \cite{kippen03}, 
\cite{kippen04}; Barraud et al. \cite{barraud03}; Sakamoto et al. \cite{sakamoto05}). 
The other spectral parameters of the prompt emission 
and the temporal behaviour of XRF afterglows at various wavelengths 
are identical to those of classical GRBs (e.g. D'Alessio, Piro \& Rossi \cite{d'alessio06}).\\
In our complete sample, we have 22 bursts classified as XRFs. In the following section 
we discuss some results obtained with this sample.
\subsection{The redshift distribution of HETE-2-XRFs}
\begin{figure}[ht]
\begin{center}
\includegraphics[width=1\linewidth,angle=0]{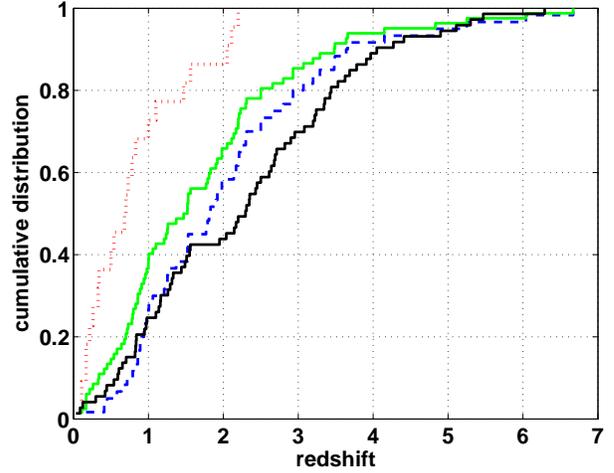}
\end{center}
\caption{\label{zdist} Four distance-scale cumulative distribution functions (cdf).
The cdf of the HETE-2 complete sample is shown with the green solid line. The two cumulative 
distributions for the `classical' GRBs and the XRFs are shown with the blue dashed line 
and the red dotted line, respectively. We note that most of the 22 XRFs 
are definitely situated at intermediate or low redshift. For comparison, we also considered all 
the long-duration bursts with redshifts detected by Swift-BAT from 
the beginning of the mission to GRB~080210 (the last Swift burst with a spectroscopic 
redshift measured to date). The corresponding cdf is the black solid line.}
\end{figure}

To explain the origin of the XRFs, Heise et al. (\cite{heise01}) proposed 
that they are {\em classical GRBs} occuring at high redshift. 
If we consider the redshift distribution of the HETE-2 complete sample, 
we note that the sub-sample of XRFs has a mean distance scale 
$<z_{sample}>_\mathrm{XRFs}^\mathrm{HETE-2}=$~0.82, lower than 
the mean redshift of the `classical' GRBs: $<z_{sample}>_\mathrm{GRBs}^\mathrm{HETE-2}=$~2.09 
(see Fig.~\ref{zdist}). The mean redshift of the XRFs contained in our sample 
is also lower than the mean redshift of the Swift XRFs derived 
by Gendre, Galli \& Piro (\cite{gendre07}): $<z>_\mathrm{XRFs}^\mathrm{Swift}=$~1.40 
for a sample of 9 XRFs with known redshifts. Consequently, we reject this hypothesis, 
as was previously done in other studies, e.g. by Barraud et al. (\cite{barraud03}) 
who highlighted this fact with the similar duration distributions they obtained 
for the XRFs and the long classical GRBs.
\subsection{Do X-ray flashes form a distinct population ?}
We consider in this paragraph the question posed, e.g., in Stratta et al. (\cite{stratta07}): 
{\it do the XRFs form a continuum extending the `classical GRBs' 
to soft energies, or do they form a distinct population, the intrinsic XRFs (i-XRFs)?}\\
Fig.~\ref{correl} (panel~1) shows the Amati relation for the complete sample 
of 82 HETE-2 bursts and Fig.~\ref{correl} (panel~2) plots the \Epi\ 
versus the distance-scales of the sources. We do not find any clear evidence 
for a distinct population and the results indicate a continuum between the classical GRBs 
and the XRFs. Indeed, according to the HETE-2 sample, several XRFs 
are situated in the gap between the group of intrinsic classical GRBs and the two intrinsic XRFs 
020903 (Sakamoto et al. \cite{sakamoto04}) and 060218 (Amati et al. \cite{amati07}), 
as is XRF~050416, a soft burst detected by Swift (Sakamoto et al. \cite{sakamoto06a}, \cite{sakamoto06b}). 
However, half of the sub-sample of the observed XRFs are part of the cluster of intrinsic 
classical GRBs when considered in their source frame. This fact was also noted 
by Stratta et al. (\cite{stratta07}). 
%
%
%
\section{The rate of GRBs in the Local Universe}
\label{s9}
\subsection{The rate of GRBs in the Local Universe, as measured by HETE-2} 
\label{rate}
The various corrections (distance scale, number of GRB within the visibility volume) 
applied to the distributions obtained in the observer frame 
allowed us to obtain the {\em unbiased} distributions of the intrinsic parameters, 
i.e. distributions that reflect more closely the true distributions of these parameters 
and the dominant characteristics of the gamma-ray burst population.\\ 
This study also allows us to derive the rate of GRBs detected 
and localized by HETE-2 ($R_0^\mathrm{H2}$). 
For that purpose, we consider that each GRB in our sample contributes 
to the local rate in proportion to:
\begin{equation}
h_b=\frac{N_{Vloc}(z=0.1)}{N_{Vmax}(z=z_{b,max})}
\end{equation} 
with the $N_V(z)$ computed according to Eq.~\ref{NV}, 
and we obtain the rate of HETE-2 GRBs during the mission,  
\begin{equation}
\tau=\frac{1}{V_{loc}}\sum_{b=1}^{n_{burst}}h_b
\end{equation}
which is 3.30~Gpc$^{-3}$.\\
\\
In order to normalize this rate per year, 
we took into account the effective monitoring time of the WXM, obtained from:
\begin{equation}
T_{m} = \frac{T_\mathrm{mission} \times T_\mathrm{\epsilon}}{4\pi} \times S_\mathrm{cov}
\end{equation}
where $T_m$ is the effective monitoring time of the WXM, $T_{mission}=$~69~months 
is the duration of the mission, $T_{\epsilon}=$~37\% is the mean observation 
efficiency during $T_{mission}$, and $S_{cov}=2\pi(1-cos~45\degr)=1.84$~sr 
the sky-coverage of the WXM\footnote{Recall that throughout this study, 
we only consider the GRBs localized by the WXM.}.\\
The effective monitoring time of the WXM is hence $T_{m}=$~0.31~yr. 
Using this, the rate of GRBs in the Local Universe per Gpc$^3$ and per year can be found 
to be $\sim$10.6~Gpc$^{-3}$yr$^{-1}$. This result is a lower limit 
because as shown in Section~\ref{s4}, the bursts that dominate 
the overall population are the ones with low $E_{peak}^{intr}$, 
and as we showed in Section~\ref{s7}, HETE-2 missed 
the bursts with intrinsic \Ep\ lower than 1 or 2~keV as well as bursts 
occuring at very high redshifts.\\
If we consider the unbiased distributions obtained in case~2 
(shown as dotted blue lines throughout the different 
Figures), we obtain 1.02~Gpc$^{-3}$, corresponding 
to a rate of $\sim$3.3~Gpc$^{-3}$yr$^{-1}$.\\
As shown in Section~\ref{s4}, the population of gamma-ray bursts 
is dominated by the X-ray flashes. This is understandable as 
XRFs are soft but also faint in the observer frame, according to 
the hardness-intensity relation derived by Barraud et al. (\cite{barraud03}). 
Therefore, if the rate of detected bursts is in fact higher 
for classical GRBs than for XRFs, we can guess that this is only due to 
instrumental limitations, such as the size and the sensitivity of the detectors. 
\subsection{Comparison with the rates of GRBs obtained in previous studies}
\label{comparison}
Several studies have derived local rates of GRBs using various methods 
mainly based on theoretical considerations.\\ 
According to the results obtained so far, the predicted 
rate of GRBs in the Universe seems to be between unity (or less) GRB 
and several hundreds per Gpc$^3$ and per year. These results seem to be related 
to the nature of the bursts considered in the studies, 
i.e (1) `classical GRBs' also called `high-luminosity GRBs' (HL-GRBs) and/or 
(2) sub-luminous (low-luminosity) GRBs (LL-GRBs).\\
\\
(1) Several authors obtained a GRB rate close to 1~Gpc$^{-3}$~yr$^{-1}$. 
The method they used is generally based on the determination 
of the luminosity function (LF) by fitting a phenomenological model 
to the $log(N)-log(P)$ relation of well-known GRB catalogs. 
Assuming that GRBs trace the star formation rate (SFR), a model of the SFR 
is adopted for it and the rate of GRBs is determined.\\
For instance, Guetta et al. (\cite{guetta04}, \cite{guetta05}) use 
the 2204 GRBs of the {\em GUSBAD Catalog} (Schmidt \cite{schmidt04}) 
and a sample consisting of 595 long GRBs detected by BATSE.
Following Schmidt (\cite{schmidt01}), they use these samples to find 
the luminosity function by fitting a single power-law model to the BATSE 
observed peak flux distribution. In addition, they assumed that the GRB rate 
follows the SFR and hence adopted either the SFR model SFR$_2$ of Porciani \& Madau 
(\cite{porciani01}) or the SFR model of Rowan-Robinson (\cite{rowan99}). 
The estimated rate lies between 0.1 and 1.1~Gpc$^{-3}$yr$^{-1}$. 
This rate is in agreement with the one found by Schmidt (\cite{schmidt01}): 
0.5~Gpc$^{-3}$yr$^{-1}$ with 1391 GRBs contained in the {\em BATSE DISCLA 2 Catalog} 
(Schmidt \cite{schmidt99}) and the one recently found by Liang et al. (\cite{liang07}) 
(1.12$_{-0.20}^{+0.43}$~Gpc$^{-3}$yr$^{-1}$) with a sample of 45 HL Swift-GRBs. 
The rate of HL HETE-2 GRBs is also consistent with all these results, 
since we obtain a lower limit $R_0^\mathrm{H2}\ga$0.45~Gpc$^{-3}$yr$^{-1}$ 
(see Part~\ref{rate_H2}).\\
\\
(2) GRB~060218, detected by Swift-BAT (Cusumano et al. \cite{cusumano06}), 
was found to be at very low-redshift $z=$0.0331 (Mirabal et al. \cite{mirabal06a}). 
Moreover, its association with the supernova SN~2006aj 
(Pian et al. \cite{pian06}) recalled the characteristics of GRB~980425, 
the closest GRB ever identified, at a redshift $z=$0.0085 
(Tinney et al. \cite{tinney98}). These GRBs associated with SNe are peculiar 
in the sense that they constitute the two nearest long duration bursts detected in 
eight years. Liang et al. (\cite{liang07}) show that the detection of these 
two events implies a local rate of low-luminosity GRBs (LL-GRBs) of 
R$_{LL}\sim$ 800~Gpc$^{-3}$yr$^{-1}$, in agreement with Guetta \& Della Valle 
(\cite{guetta07}) who find a rate of GRB 980425-like events of 
$\sim\,430$~Gpc$^{-3}$yr$^{-1}$ and a rate of GRB 060218-like events of 
$\sim\,350$~Gpc$^{-3}$yr$^{-1}$.\\
The rate of LL-GRBs, which is considerably higher than the rate of HL-GRBs, 
could be explained in two ways: first, the $log(N)-log(P)$ distribution 
of BATSE bursts often used to fit the GRB luminosity function cannot constrain 
it down to such low luminosities, because a GRB~060218-like event could not 
trigger BATSE. Second, the use of a phenomenological model for the GRB LF 
requires a choice for the extrema values $L_{min}$ and $L_{max}$. 
However, due to instrumental limitations, the true $L_{min}$ of GRBs 
is unknown. Thus, the number of low-luminosity GRB is quite uncertain.
\subsection{Discussion}
\label{rate_H2}
We showed that the study of the complete sample of GRBs localized 
by the WXM on-board HETE-2 led to a rate $R_0^\mathrm{H2}\ga$~10.6~Gpc$^{-3}$yr$^{-1}$. 
Our result is independent of any assumption on the luminosity function, 
as it is based on measured data. 
We note that this measured rate is at least 10 times higher 
than the HL-GRB rate and 10 times lower than the LL-GRB one. 
{\em How can we explain this `intermediate' result?} 
\begin{itemize}
\item ($R_0^\mathrm{H2}< R_0^\mathrm{LL}$): 
We first note that we have no event like GRB~980425 
or GRB~060218 in our sample. 
Previous studies showed that it was impossible for an HETE-like 
mission with a WXM-like camera to detect a burst like XRF~060218. 
This burst occured at a distance $d=$145~Mpc, and Soderberg et al. (\cite{soderberg06d}) 
show that the WXM could have detected it at a maximum distance 
of 110~Mpc. For a burst similar to GRB~980425 occurring at $d=$~36.1~Mpc, 
the WXM could have detected it to a distance $d=$~60~Mpc. However, Guetta et al. 
(\cite{guetta04}) show that with their model of the LF, the local rate 
of nearby bright bursts observable by HETE-2 within this distance 
is only 0.057~yr$^{-1}$.\\
\\
\item ($R_0^\mathrm{H2}> R_{0}^\mathrm{HL}$): 
If we now imagine that HETE-2 was a mission that could not detect 
X-ray flashes, we can try to re-do all the steps to derive the rate 
of GRBs detected and localized by it (see Part. \ref{rate}) 
in the sample of 60 `classical' GRBs. We obtain $R_0^\mathrm{H2}\ga$~0.45~Gpc$^{-3}$yr$^{-1}$. 
This result is in good agreement with the rates usually obtained 
when only the HL-GRBS are considered: $\sim$[0.1--1]~Gpc$^{-3}$yr$^{-1}$.
\end{itemize}
Consequently, we infer that the difference between the rate of HL-GRBs obtained with BATSE 
and the rate measured with HETE-2 is due to the X-ray flashes. 
These soft events cannot be detected in the BATSE energy range (50-300~keV). 
With our method based on observed properties of GRBs, 
XRFs and classical GRBs can be distinguished. The average lower redshift of XRFs 
leads to higher weights for the XRFs than for classical GRBs\footnote{Recall that 
the weight taken for each burst is inversely proportional to the number of GRBs 
within its visibility volume and is obviously redshift-dependent.}. 
In fact, it appears that XRFs constitute 
an intermediate class in terms of frequency between the 
HL-GRBs and the LL-GRBs, and the rate of XRFs could be higher 
than the rate of  classical GRBs by a factor of $\sim$10--100.\\
\\ 
Despite their higher softness, faintness and frequency, 
our study brings the XRFs closer to the classical GRBs. This is not the case 
for GRB~980425 or XRF~060218-like events that are undetectable 
and out of reach of a HETE-2-like mission. 
\subsection{Comparison with the rate of Type Ib/c SNe}
If we compare the measured rate of GRBs based on the complete HETE-2 sample and the rate 
of SNe~Ib/c considered in Soderberg et al. (\cite{soderberg06d}), 
$R_0^\mathrm{SNeIb/c}=$~9$_{-5}^{+3}~\times$~10$^3$~Gpc$^{-3}$yr$^{-1}$ 
measured by Cappellaro, Evans \& Turatto (\cite{cappellaro99}) and Dahlen et al. (\cite{dahlen04}), 
the ratio is $\gtrsim$0.1\%.\\
This low ratio highlights the fact that not all the type Ib/c SNe produce GRBs. 
The main explanation invokes a mildly relativistic jet, 
which is the source of the GRB, but that does not appear in all SNe~Ib/c. 
Indeed, if the optical luminosities of GRBs and SNe can be considered to be similar, 
optical spectroscopy reveals broad absorption lines in SNe spectra associated with GRBs, 
indicative of fast ejecta. These peculiar highly energetic SNe, often designed 
{\em Hypernovae} ({\em HNe}) would not represent more than 3\% of the SNe~Ib/c, 
as obtained by the studies based on the comparison of radio luminosities of GRB 
afterglows and SNe~Ib/c (see e.g. Berger et al. \cite{berger03}; 
Soderberg et al. \cite{soderberg06b}; Soderberg \cite{soderberg07}).\\
From previous studies of the predicted local rate of GRBs, and with our measured 
local rate, we now have a ratio between the SNe~Ib/c and the GRBs lying in 
the range $\sim$[0.1\%-3\%]. This small ratio could imply that the escape 
of the ultra-relativistic jet from the progenitor star is not the only ingredient 
needed to produce a GRB. In addition, several studies of stellar rotation, 
binarity, asymmetry and metallicity have shown that the progenitors must have 
some special characteristics in order to produce a gamma-ray burst.
%
%
%
\section{Summary}
\label{s10}
In this paper we have presented the study of a sample 
containing {\em all} the gamma-ray bursts 
detected by HETE-2 above a certain threshold {\em and} localized by the WXM. 
Our main goal was (1) to provide intrinsic distributions 
of the main global properties associated with the GRBs, i.e. taking into account 
the cosmological effects. The parameters studied are the peak energy (\Ep), 
the duration ($T_{90}$) and the total energy (\Eiso). The second purpose 
of this paper is (2) to derive unbiased distributions 
of these properties, taking into account the spatial density of GRBs, 
and to measure the rate of HETE-2 GRBs in the local universe.\\
\\ 
(1) The main conclusions are the following: 
\begin{itemize}
\item the intrinsic \Ep\ distribution is quite broad, containing at 
the extreme sides of the energy range a large sample of X-ray flashes 
and few `classical' hard bursts. Nevertheless, we caution 
that the true width of this distribution is probably even broader 
than derived here, since the detector characteristics prevented HETE-2 
from detecting bursts with an intrinsic peak energy 
of 1 or 2~keV or, at the opposite end, bursts with an intrinsic peak energy of a few MeV.    
\item the intrinsic $T_{90}$ distribution shows that HETE-2 did not localize 
very long duration bursts ($T_{90}>$1000~s). This may be a limitation 
of the orbit because due to its antisolar position, the detectors are turned off 
every half-orbit when they have the Earth in their field of view, 
i.e. every $\sim\,45$~min. On the other hand, some events have an intrinsic 
duration below 2~seconds and might be considered as `intrinsically short'. 
\item the intrinsic \Eiso\ distribution extends over more 
than four decades in energy. Most of the observed bursts have \Eiso\ higher 
than 10$^{52}$~erg: the `classical' GRBs. The distribution extends down to 
lower \Eiso\ (10$^{50}$--10$^{51}$~erg), corresponding to the sample of HETE-2 XRFs.
\end{itemize}
(2) The main conclusions are the following:
\begin{itemize}
\item the `unbiased' distributions are dramatically different from their uncorrected shapes. 
Indeed, the unbiased \Ep\ distribution shows a predominence of bursts 
with intrinsic \Ep\ lower than 10~keV and the unbiased \Eiso\ distribution 
shows a predominence of bursts with \Eiso\ lower than 10$^{50}$--10$^{51}$~erg. 
The \Eiso\ distribution can be fitted between 10$^{50}$ and 10$^{54}$~erg 
by a simple power-law with a slope of -0.84.\\
However, we caution that the `simple' distributions of the intrinsic parameters 
probably suffer from biases. The same point may be made 
for the shapes of the 'unbiased' distributions: 
the true ones are probably also different, as shown by the 
two studies performed using different distances for the bursts 
without secure redshifts -- case~1: pseudo-redshifts 
and case~2: known spectroscopic redshifts 
randomly attributed -- which lead to different shapes 
for the \Ep\ and the $T_{90}$ distributions. This emphasizes 
the predominence of X-ray flashes and the existence of a typical intrinsic 
duration for the GRBs in the first case, whereas a nearly constant number 
of GRBs through both the \Ep\ energy range and the durations is highlighted 
in the second case. We may guess that the true distributions are situated 
between the results of these two cases. 
\item a possible application of our study is the determination 
of the rate of GRBs measured by HETE-2 in the Local Universe. 
We obtained a lower limit of 10.6~Gpc$^{-3}$yr$^{-1}$, which is intermediate 
in magnitude between the rates usually found when a population of 
High-Luminosity GRBs is considered ($\sim$1~Gpc$^{-3}$yr$^{-1}$) 
or when the Low-Luminosity GRBs are also taken into account  ($\ga$100~Gpc$^{-3}$yr$^{-1}$). 
We explain this result with two arguments. First, HETE-2 did not detect LL-GRBs 
like GRB~980425 or XRF~060218 -- mainly due to the small size 
of its detectors -- so the rate cannot be as high as the one found 
for LL-GRBs. Second, it is the sample of XRFs detected by HETE-2 
that makes the difference between the rate we obtained and the one 
found for HL-GRBs. As a consequence, thanks to the construction of unbiased distributions 
of the intrinsic properties of HETE-2 GRBS, we experimentally showed 
in this paper the dominance in terms of frequency of the X-ray flashes 
over the `classical GRBs', a frequency which is nevertheless 
lower than the sub-luminous bursts, and much smaller than the Type Ib/c supernovae.
\end{itemize}  
%
%
%
\begin{acknowledgements}
The HETE-2 mission was supported in the US by NASA contract NASW-4690, 
in France by CNES contract 793-01-8479 and in Japan in part 
by the Ministry of Education, Culture, Sports, Science and Technology 
and by Grant-in-Aid for Scientific Research on Priority Areas 19047001. 
The authors acknowledge the valuable support of the HETE-2 
Operation Team. AP is supported in France by the Ministry of National Education, 
Research and Technology. YEN is supported by the JSPS Research Fellowships 
for Young Scientists. This work is supported in part by a special postdoctoral 
researchers program in RIKEN. KH is grateful for support under MIT Contract 
SC-R-293291. G.~Pizzichini aknowledges financial support as part 
of ASI contract I/088/06/0. Finally, the authors acknowledge the referee for 
his/her valuable and relevant remarks that helped improving the content of 
this paper. 
\end{acknowledgements}
%
%

%
%
%
\longtab{2}{
\begin{longtable}[ht]{l c c c c c c c l}
\caption{\label{tab1} Observed properties and distances used in this study.\\
The GRB names in the format {\em YYMMDD} are given in column~1. Columns~2 to 5 
contain the spectral parameters measured for each burst: the slope 
of the power-laws at low-energy ($\alpha$) and at high-energy when 
possible ($\beta$); the peak energy (\Epobs) in keV and the fluence 
in the energy range 2-30~keV ($S_X$), in units of $10^{-7}$~ergs~cm$^{-2}$. 
The $T_{90}$ durations in the FREGATE band $B$ (6-80~keV) 
are reported in column~6. The signal-to-noise ratio (SNR) corresponding 
to the best combination of energy range and time resolution 
(see Section \ref{s2}) are in column~7. The \zsamp\ 
(i.e. the spectroscopic/photometric redshifts when available 
or the pseudo-redshifts otherwise) are reported in column~8. 
The last column contains the references for the redshifts.}\\
\hline\hline
GRB & $\alpha$ & $\beta$ & $E_{peak}^{obs}$~(keV) & $S_{X}$ & $T_{90,B}$ & SNR & \zsamp & Ref.~\zsamp\\
\hline
\endfirsthead
\caption{(continued)}\\
\hline\hline
GRB & $\alpha$ & $\beta$ & $E_{peak}^{obs}$~(keV) & $S_{X}$ & $T_{90,B}$ & SNR & \zsamp & Ref.~\zsamp\\
\hline
\endhead
\hline
\endfoot
001226&$-1$&$-2.094_{-0.293}^{+0.194}$&$22.32_{-7.42}^{+7.48}$&$17.03_{-1.7}^{+1.7}$&$90.69\pm2.5$&$10.35$&$0.98_{-0.3}^{+0.3}$&$[0]$\\
010213&$-1$&$-3_{-0.5}^{+0.2}$&$3.4_{-0.4}^{+0.4}$&$7.9_{-0.5}^{+0.3}$&$20.12\pm2.16$&$41.98$&$0.17_{-0.06}^{+0.06}$&$[0]$\\
010225&$-1.3_{-0.3}^{+0.3}$&$--$&$32_{-9}^{+27}$&$3.5_{-0.4}^{+0.4}$&$6.3\pm0.6$&$7.94$&$1.47_{-0.55}^{+0.55}$&$[0]$\\
010326&$-1.08_{-0.22}^{+0.25}$&$--$&$51.8_{-11.3}^{+18.6}$&$2.42_{-0.39}^{+0.39}$&$2.15\pm0.32$&$20.71$&$2.93_{-1.65}^{+1.65}$&$[0]$\\
010612&$-1.1_{-0.2}^{+0.2}$&$--$&$240_{-82}^{+290}$&$8.8_{-1.3}^{+1.4}$&$49.98\pm3.54$&$23.52$&$5.25_{-2.2}^{+2.2}$&$[0]$\\
010613&$-1_{-0.3}^{+0.3}$&$-2_{-0.2}^{+0.1}$&$46_{-10}^{+18}$&$102_{-7}^{+7}$&$146.74\pm2.44$&$45.05$&$0.64_{-0.2}^{+0.2}$&$[0]$\\
010629&$-1.1_{-0.1}^{+0.1}$&$--$&$46_{-4}^{+5}$&$25.4_{-1.7}^{+1.7}$&$15.8\pm0.28$&$40.15$&$0.91_{-0.4}^{+0.4}$&$[0]$\\
010921&$-1.6_{-0.1}^{+0.1}$&$--$&$89_{-14}^{+22}$&$72_{-3}^{+3}$&$21.92\pm1.18$&$71.52$&$0.45_{-0.005}^{+0.005}$&$[1;2]$\\
010928&$-0.7_{-0.1}^{+0.1}$&$--$&$410_{-75}^{+120}$&$13.7_{-1.4}^{+1.2}$&$31.73\pm0.97$&$39.04$&$3.64_{-0.8}^{+0.8}$&$[0]$\\
011019&$-1.4$&$--$&$19_{-9}^{+18}$&$3_{-0.6}^{+0.6}$&$22.5\pm2.24$&$8.75$&$0.5_{-0.5}^{+0.5}$&$[0]$\\
011130&$--$&$-2.7_{-0.3}^{+0.3}$&$<~3.9$&$5.9_{-1}^{+1}$&$10.64\pm2.86$&$7.66$&$<~0.17$&$[0]$\\
011212&$-1.23$&$-2.15_{-1.33}^{+0.26}$&$<~3.74$&$4.85_{-0.49}^{+0.49}$&$39.62\pm6.55$&$8.09$&$0.34_{-0.22}^{+0.22}$&$[0]$\\
020124&$-0.87_{-0.16}^{+0.19}$&$-2.6_{-0.65}^{+0}$&$82_{-17}^{+17}$&$20_{-1.4}^{+1.4}$&$51.17\pm1.55$&$21.63$&$3.198_{-0.004}^{+0.004}$&$[3]$\\
020127&$-1_{-0.1}^{+0.1}$&$--$&$100_{-24}^{+47}$&$6.7_{-0.5}^{+0.5}$&$6.99\pm0.26$&$27.65$&$1.9_{-0.4}^{+0.2}$&$[4]$\\
020305&$-1.06_{-0.13}^{+0.14}$&$-2.3$&$245.1_{-99.9}^{+276.7}$&$27.49_{-4.82}^{+1.94}$&$39.06\pm1.17$&$28.54$&$1.98_{-0.7}^{+0.7}$&$[0]$\\
020317&$-0.6_{-0.5}^{+0.6}$&$--$&$28_{-7}^{+13}$&$2.2_{-0.4}^{+0.4}$&$7.14\pm1.04$&$12.11$&$2.11_{-0.55}^{+0.55}$&$[0]$\\
020331&$-0.8_{-0.1}^{+0.1}$&$--$&$92_{-14}^{+21}$&$16_{-1}^{+1}$&$179.4\pm5.99$&$28.15$&$2.21_{-0.7}^{+0.7}$&$[0]$\\
020625&$-1.1$&$--$&$8.5_{-2.9}^{+5.4}$&$2.4_{-0.5}^{+0.6}$&$13.98\pm3.82$&$8.18$&$0.73_{-0.13}^{+0.13}$&$[0]$\\
020801&$-0.3_{-0.3}^{+0.4}$&$-2_{-0.3}^{+0.2}$&$53_{-11}^{+14}$&$26_{-3}^{+3}$&$460.33\pm3.63$&$15.54$&$1.21_{-1}^{+1}$&$[0]$\\
020812&$-1.1_{-0.3}^{+0.3}$&$--$&$88_{-30}^{+110}$&$8_{-1}^{+1}$&$18.28\pm1.55$&$14.63$&$3.48_{-1.8}^{+1.8}$&$[0]$\\
020813&$-1.36_{-0.05}^{+0.05}$&$-2.3$&$253.4_{-35.6}^{+52.4}$&$147.91_{-7}^{+7.32}$&$87.34\pm0.6$&$92.98$&$1.255_{-0.005}^{+0.005}$&$[5]$\\
020819&$-0.9_{-0.1}^{+0.2}$&$-2_{-0.5}^{+0.2}$&$50_{-12}^{+18}$&$25.2_{-1.1}^{+1.1}$&$28.8\pm7.19$&$50.61$&$0.41_{-0.01}^{+0.01}$&$[6]$\\
021004&$-1_{-0.2}^{+0.2}$&$--$&$80_{-23}^{+53}$&$7.7_{-0.7}^{+0.7}$&$48.94\pm2.5$&$13.85$&$2.31_{-0.01}^{+0.01}$&$[7;8;9]$\\
021016&$-1.2_{-0.12}^{+0.11}$&$-2.3$&$226.2_{-86.4}^{+268.7}$&$34.17_{-3.59}^{+2.44}$&$80.63\pm0.95$&$28.55$&$2.8_{-1.6}^{+1.6}$&$[0]$\\
021021&$-1.3$&$--$&$15_{-8}^{+14}$&$2.5_{-0.6}^{+0.6}$&$18.9\pm5.13$&$6.28$&$1_{-0.5}^{+0.5}$&$[0]$\\
021104&$-1.1_{-0.5}^{+0.5}$&$--$&$28_{-8}^{+17}$&$10_{-2}^{+2}$&$21.18\pm4.08$&$10.52$&$1.1_{-0.3}^{+0.3}$&$[0]$\\
021112&$-0.9_{-0.3}^{+0.4}$&$--$&$57_{-21}^{+39}$&$1.3_{-0.3}^{+0.3}$&$3.24\pm1.28$&$9.86$&$4.15_{-1.9}^{+1.9}$&$[0]$\\
021211&$-0.805_{-0.105}^{+0.112}$&$-2.37_{-0.31}^{+0.18}$&$46.8_{-5.1}^{+5.8}$&$13.6_{-0.5}^{+0.5}$&$4.23\pm0.27$&$76.24$&$1.006_{-0.001}^{+0.001}$&$[10;11]$\\
030115&$-1.3_{-0.1}^{+0.1}$&$--$&$83_{-22}^{+53}$&$7.9_{-0.6}^{+0.6}$&$20.33\pm3.54$&$27.29$&$2.2_{-0.1}^{+0.1}$&$[12]$\\
030226&$-0.9_{-0.2}^{+0.2}$&$--$&$97_{-17}^{+27}$&$13_{-1}^{+1}$&$76.23\pm3.96$&$16.3$&$1.98_{-0.02}^{+0.02}$&$[13;14;15;16;17]$\\
030324&$-1.5_{-0.1}^{+0.1}$&$--$&$150_{-65}^{+630}$&$5.5_{-0.4}^{+0.4}$&$10.98\pm2.84$&$21.27$&$>~2.3$&$[0]$\\
030328&$-1.14_{-0.03}^{+0.03}$&$-2.1_{-0.4}^{+0.2}$&$130_{-13}^{+14}$&$82_{-1}^{+1}$&$138.27\pm3.05$&$55.68$&$1.5216_{-0.0006}^{+0.0006}$&$[18;19;20;21]$\\
030329&$-1.32_{-0.02}^{+0.02}$&$-2.44_{-0.08}^{+0.08}$&$70.2_{-2.3}^{+2.3}$&$576_{-5}^{+5}$&$25.91\pm0.39$&$581.25$&$0.1685_{-0.0001}^{+0.0001}$&$[22;23;24]$\\
030416&$--$&$-2.3_{-0.2}^{+0.1}$&$2.6_{-1.8}^{+0.5}$&$9_{-0.9}^{+0.9}$&$14.29\pm2.29$&$18.02$&$0.11_{-0.04}^{+0.04}$&$[0]$\\
030418&$-1.5_{-0.1}^{+0.1}$&$--$&$46_{-14}^{+32}$&$17.1_{-1.1}^{+1.1}$&$139.23\pm7.76$&$9.26$&$3.07_{-1.7}^{+1.7}$&$[0]$\\
030429&$-1.1_{-0.2}^{+0.3}$&$--$&$35_{-8}^{+12}$&$4.7_{-0.5}^{+0.5}$&$12.95\pm2.69$&$13.25$&$2.658_{-0.004}^{+0.004}$&$[25]$\\
030519&$-0.8_{-0.1}^{+0.1}$&$-1.7_{-0.1}^{+0.1}$&$138_{-15}^{+18}$&$87.1_{-2.4}^{+2.4}$&$12.85\pm0.56$&$193.85$&$0.86_{-0.1}^{+0.1}$&$[0]$\\
030528&$-1.3_{-0.1}^{+0.2}$&$-2.7_{-1}^{+0.3}$&$32_{-5}^{+5}$&$62_{-3}^{+3}$&$62.8\pm4.49$&$28.11$&$0.782_{-0.001}^{+0.001}$&$[26]$\\
030723&$--$&$-1.9_{-0.2}^{+0.2}$&$<~8.9$&$2.8_{-0.5}^{+0.5}$&$9.63\pm1.5$&$9.74$&$0.54_{-0.15}^{+0.15}$&$[0]$\\
030725&$-1.51_{-0.04}^{+0.04}$&$--$&$102_{-14}^{+19}$&$94_{-2}^{+2}$&$174.31\pm17.07$&$88.56$&$0.89_{-0.2}^{+0.2}$&$[0]$\\
030821&$-0.9_{-0.1}^{+0.1}$&$--$&$84_{-11}^{+15}$&$10_{-0.6}^{+0.6}$&$19.42\pm0.44$&$21.43$&$1.77_{-0.5}^{+0.5}$&$[0]$\\
030823&$-1.3_{-0.2}^{+0.2}$&$--$&$27_{-5}^{+8}$&$23.1_{-1.6}^{+1.6}$&$50.39\pm3.11$&$15.49$&$0.83_{-0.3}^{+0.3}$&$[0]$\\
030824&$--$&$-2.1_{-0.1}^{+0.1}$&$6.1_{-4.2}^{+1.9}$&$8.9_{-1.1}^{+1.1}$&$10.13\pm1.04$&$12.84$&$0.26_{-0.1}^{+0.1}$&$[0]$\\
030913&$-0.8_{-0.2}^{+0.3}$&$--$&$120_{-37}^{+110}$&$1.8_{-0.2}^{+0.2}$&$6.58\pm2.3$&$14.23$&$6.04_{-2.7}^{+2.7}$&$[0]$\\
031026&$-1.13_{-0.21}^{+0.36}$&$-2.3$&$870.3_{-679}^{+129.7}$&$3.56_{-3.46}^{+1.26}$&$65.58\pm4.55$&$8.46$&$6.67_{-2.9}^{+2.9}$&$[0]$\\
031109A&$-1.17_{-0.04}^{+0.03}$&$-2.3$&$185.2_{-23.7}^{+29.8}$&$98.3_{-2.9}^{+3}$&$57.32\pm0.46$&$78.68$&$0.94_{-0.2}^{+0.2}$&$[0]$\\
031109B&$-1.27_{-0.24}^{+0.28}$&$--$&$37.7_{-12.29}^{+28.4}$&$4.9_{-0.6}^{+0.5}$&$531.25\pm72.25$&$8.7$&$2.05_{-0.9}^{+0.9}$&$[0]$\\
031111A&$-0.82_{-0.05}^{+0.05}$&$--$&$404.4_{-51.4}^{+67.8}$&$14.9_{-0.7}^{+0.6}$&$7.94\pm0.53$&$151.81$&$2.14_{-0.4}^{+0.4}$&$[0]$\\
031111B&$--$&$-2.2_{-0.38}^{+0.2}$&$6.01_{-5.01}^{+3.8}$&$9.85_{-0.99}^{+0.99}$&$27.34\pm2.16$&$5.81$&$0.11_{-0.17}^{+0.17}$&$[0]$\\
031203&$-1.18_{-0.09}^{+0.1}$&$-2.3$&$148.2_{-23.91}^{+32.45}$&$22.68_{-1.68}^{+1.44}$&$10.38\pm0.29$&$92.35$&$2.17_{-0.7}^{+0.7}$&$[0]$\\
031220&$-1.51_{-0.2}^{+0.23}$&$-3.25_{-6.75}^{+1.3}$&$46.9_{-16.4}^{+62.1}$&$5.5_{-0.5}^{+0.6}$&$9.69\pm0.93$&$13.9$&$1.53_{-1.2}^{+1.2}$&$[0]$\\
040319&$-0.89_{-0.2}^{+0.24}$&$--$&$56.6_{-9.7}^{+14.1}$&$6_{-0.5}^{+0.6}$&$6.1\pm0.85$&$23.12$&$1.79_{-1.2}^{+1.2}$&$[0]$\\
040423&$-1.02_{-0.27}^{+0.31}$&$--$&$30.7_{-4}^{+4.8}$&$22.7_{-2.2}^{+2.6}$&$45.87\pm1.13$&$17.47$&$1.26_{-0.7}^{+0.7}$&$[0]$\\
040425&$-0.86_{-0.08}^{+0.09}$&$--$&$299.9_{-48.7}^{+72.9}$&$17.2_{-1.4}^{+1.4}$&$138.81\pm1.65$&$11.16$&$2.23_{-0.5}^{+0.5}$&$[0]$\\
040511&$-0.92_{-0.14}^{+0.15}$&$--$&$93.9_{-11.3}^{+15.7}$&$31.7_{-2.8}^{+0.3}$&$45.8\pm0.53$&$36.61$&$1.83_{-0.5}^{+0.5}$&$[0]$\\
040701&$--$&$-2.3_{-0.31}^{+0.25}$&$<~3.44$&$5.44_{-0.55}^{+0.55}$&$11.67\pm5.72$&$12.76$&$0.2146_{-0.0005}^{+0.0005}$&$[27]$\\
040709&$-1.25_{-0.07}^{+0.08}$&$--$&$70.9_{-6.7}^{+8.3}$&$80.6_{-4}^{+4.2}$&$82.48\pm0.24$&$58.3$&$1_{-0.3}^{+0.3}$&$[0]$\\
040802&$-0.85_{-0.2}^{+0.23}$&$--$&$92.2_{-13}^{+18.8}$&$4.79_{-0.68}^{+0.68}$&$3.05\pm0.18$&$39.39$&$1.81_{-0.65}^{+0.65}$&$[0]$\\
040825A&$-1.51_{-0.22}^{+0.22}$&$--$&$60_{-30}^{+30}$&$9.6_{-1.5}^{+1.5}$&$39.2\pm2.01$&$10.47$&$1.06_{-0.5}^{+0.5}$&$[0]$\\
040825B&$-1.48_{-0.17}^{+0.21}$&$--$&$25.1_{-8.2}^{+15.6}$&$12_{-0.9}^{+0.9}$&$15.8\pm5.45$&$10.3$&$2.2_{-0.8}^{+0.8}$&$[0]$\\
040912A&$-1.17_{-0.62}^{+1.54}$&$-2.49_{-0.4}^{+0.24}$&$14.16_{-4.32}^{+3.32}$&$20.36_{-2.04}^{+2.04}$&$9.21\pm0.27$&$39.44$&$0.33_{-0.3}^{+0.3}$&$[0]$\\
040912B&$-1.25_{-0.58}^{+0.82}$&$--$&$17_{-13}^{+13}$&$9.5_{-0.95}^{+0.95}$&$122.41\pm7.18$&$8.19$&$1.563_{-0.001}^{+0.001}$&$[28]$\\
040916&$-1$&$--$&$<~3.5$&$7.74_{-0.91}^{+0.91}$&$349\pm10.23$&$6.66$&$<~0.7$&$[0]$\\
040924&$-1.03_{-0.09}^{+0.09}$&$--$&$41.1_{-2.3}^{+2.3}$&$23.4_{-0.24}^{+0.24}$&$3.37\pm0.08$&$85.97$&$0.859_{-0.005}^{+0.005}$&$[29]$\\
041004&$-1.3_{-0.17}^{+1.17}$&$--$&$53.7_{-7.5}^{+11.7}$&$133.8_{-18.4}^{+19.7}$&$50.13\pm2.49$&$35.64$&$0.58_{-0.2}^{+0.2}$&$[0]$\\
041006&$-1.3_{-0.05}^{+0.05}$&$--$&$47.7_{-2.7}^{+3}$&$38.9_{-0.9}^{+0.6}$&$22.08\pm0.33$&$120.37$&$0.716_{-0.005}^{+0.005}$&$[30]$\\
041016&$-1.13_{-0.21}^{+0.36}$&$-2.3$&$165.3_{-75.2}^{+731.8}$&$8.21_{-1.93}^{+1.53}$&$21.96\pm1.7$&$13.59$&$3.49_{-1.7}^{+1.7}$&$[0]$\\
041127&$-1.01_{-0.16}^{+0.27}$&$-2.27_{-0.47}^{+0.24}$&$35_{-10.2}^{+10}$&$26.9_{-1.4}^{+0.07}$&$49.14\pm2.26$&$25.19$&$0.96_{-0.5}^{+0.5}$&$[0]$\\
041211&$-0.69_{-0.2}^{+0.48}$&$-1.48_{-0.54}^{+0.16}$&$132_{-76.3}^{+116}$&$10.6_{-0.8}^{+0.4}$&$113.84\pm4.2$&$44.38$&$3.29_{-0.9}^{+0.9}$&$[0]$\\
050123&$-0.71_{-0.2}^{+0.22}$&$--$&$40.2_{-5}^{+6}$&$6.7_{-3.7}^{+0.5}$&$14.72\pm1.17$&$25.06$&$1.53_{-0.85}^{+0.85}$&$[0]$\\
050209&$-1.52_{-0.15}^{+0.3}$&$-2.3$&$445_{-350.5}^{+555}$&$5.5_{-5.4}^{+0.82}$&$18.38\pm0.59$&$15.37$&$2.93_{-1.6}^{+1.6}$&$[0]$\\
050408&$-1.76_{-0.06}^{+0.09}$&$-2.2_{-0.28}^{+0.16}$&$25.9_{-7.3}^{+10.5}$&$32.4_{-1.1}^{+0.3}$&$28.39\pm0.56$&$50.85$&$1.2357_{-0.0002}^{+0.0002}$&$[31;32]$\\
050509&$--$&$--$&$<~19$&$6_{-0.6}^{+0.6}$&$21\pm5$&$7.06$&$0.68_{-0.3}^{+0.3}$&$[0]$\\
050729&$-0.61_{-0.32}^{+0.43}$&$-2.3$&$78.42_{-16.03}^{+23.22}$&$3.97_{-0.65}^{+0.69}$&$5.23\pm0.51$&$24.35$&$2.5_{-0.75}^{+0.75}$&$[0]$\\
050807&$-1.53_{-0.22}^{+0.42}$&$-2.3$&$69.47_{-31.12}^{+102.9}$&$14.15_{-3.36}^{+1.84}$&$10.32\pm1$&$19.17$&$0.79_{-0.75}^{+0.75}$&$[0]$\\
050922&$-0.83_{-0.26}^{+0.23}$&$--$&$130.5_{-26.8}^{+50.9}$&$5.4_{-0.54}^{+0.54}$&$6.13\pm1.4$&$38.25$&$2.198_{-0.001}^{+0.001}$&$[33]$\\
051021&$-1.04_{-0.1}^{+0.11}$&$--$&$96_{-16.6}^{+26}$&$19.2_{-0.7}^{+0.5}$&$36.79\pm1.13$&$35.54$&$1.37_{-0.5}^{+0.5}$&$[0]$\\
051022&$-1.01_{-0.03}^{+0.02}$&$-1.95_{-0.14}^{+0.25}$&$213_{-18}^{+18}$&$214_{-2}^{+2}$&$178\pm8$&$226.38$&$0.8_{-0.01}^{+0.01}$&$[34]$\\
051028&$-0.94_{-0.16}^{+0.18}$&$--$&$249_{-70}^{+167}$&$7.99_{-0.67}^{+0.59}$&$14.66\pm1$&$25.35$&$3.66_{-1.8}^{+1.8}$&$[0]$\\
051211&$-0.07_{-0.41}^{+0.5}$&$--$&$121_{-20.3}^{+33}$&$0.765_{-0.228}^{+0.228}$&$4.9\pm0.61$&$18.83$&$4.83_{-1.9}^{+1.9}$&$[0]$\\
060115&$-0.74_{-0.1}^{+0.12}$&$-1.81_{-0.21}^{+0.12}$&$94.6_{-17.6}^{+21.4}$&$22.4_{-0.5}^{+0.4}$&$20.68\pm0.45$&$87.09$&$1.52_{-0.25}^{+0.25}$&$[0]$\\
060121&$-0.79_{-0.11}^{+0.12}$&$--$&$114_{-10.9}^{+14.2}$&$8.98_{-0.67}^{+0.67}$&$2.61\pm0.1$&$80.37$&$1.92_{-0.35}^{+0.35}$&$[0]$\\
\hline
\end{longtable}
\noindent
{\small [0]= our work; [1]= Djorgovski (\cite{djorgovski01}); [2]= Price et al. (\cite{price02}); 
[3]= Hjorth et al. (\cite{hjorth03b}); [4]= Berger et al. (\cite{berger07}); 
[5]= Barth et al. (\cite{barth03}); [6]= Jakobsson et al. (\cite{jakobsson05a}); 
[7]= Savaglio et al. (\cite{savaglio02}); [8]= Castro-Tirado et al. (\cite{castro-tirado02}); 
[9]= Lazzati et al. (\cite{lazzati06}); [10]= Vreeswijk et al. (\cite{vreeswijk02}); 
[11]= Fox et al. (\cite{fox03}); [12]= Levan et al. (\cite{levan06}); 
[13]= Ando et al. (\cite{ando03a}); [14]= Ando et al. (\cite{ando03b}); 
[15]= Greiner et al. (\cite{greiner03a}); [16]= Price et al. (\cite{price03}); 
[17]= Shin et al. (\cite{shin06}); 
[18]= Martini et al. (\cite{martini03}); [19]= Rol et al. (\cite{rol03}); 
[20]= Fugazza et al. (\cite{fugazza03}); [21]= Maiorano et al. (\cite{maiorano06}); 
[22]= Greiner et al. (\cite{greiner03b}); [23]= Caldwell et al. (\cite{caldwell03}); 
[24]= Bloom et al. (\cite{bloom03}); [25]= Jakobsson et al. (\cite{jakobsson04}); 
[26]= Rau, Salvato \& Greiner (\cite{rau05}); [27]= Kelson et al. (\cite{kelson04}); 
[28]= Stratta et al. (\cite{stratta07}); [29]= Wiersema et al. (\cite{wiersema04}); 
[30]= Fugazza et al. (\cite{fugazza04}); [31]= Berger, Gladders \& Oemler (\cite{berger05}); 
[32]= Prochaska et al. (\cite{prochaska05}); [33] = Jakobsson et al. (\cite{jakobsson05b}); 
[34]= Gal-Yam et al. (\cite{gal-yam05}).
}
}
%
%
\longtab{3}{
{
\begin{longtable}[ht]{l c c c c c}
\caption{\label{tab2} Intrinsic properties of our complete sample of HETE-2 localized GRBs.\\
For each burst, the values reported in columns~2, 3 and 4 are base 10 logarithms 
which enable us to obtain the individual lognormal distributions of the \Ep, $T_{90,B}$ 
and \Eiso. The $z_{max}$ used to derive the corresponding $V_{max}$ 
are shown in column~5. The resulting $V/V_{max}$ are reported in the last column.}\\
\hline\hline
GRB & log~$E_{peak}^{intr}$ & log~$T_{90,B}^{intr}$ & log~$E_{iso,52}$ & $z_{max}$ &  $V/V_{max}$\\
\hline
\endfirsthead
\caption{(continued)}\\
\hline\hline
GRB & log~$E_{peak}^{intr}$ & log~$T_{90,B}^{intr}$ & log~$E_{iso,52}$ & $z_{max}$ &  $V/V_{max}$\\
\hline
\endhead
\hline
\endfoot
001226......&$1.648\pm0.091$&$1.66\pm0.037$&$0.516\pm0.112$&$1.12\pm0.19$&$0.779\pm0.012$\\
010213......&$0.593\pm0.03$&$1.239\pm0.028$&$-2.172\pm0.023$&$0.39\pm0.08$&$0.091\pm0.009$\\
010225......&$2.003\pm0.122$&$0.403\pm0.059$&$-0.173\pm0.08$&$1.5\pm0.3$&$1$\\
010326......&$2.341\pm0.111$&$-0.264\pm0.106$&$0.518\pm0.115$&$4.85\pm1.47$&$0.552\pm0.049$\\
010612......&$3.306\pm0.168$&$0.916\pm0.095$&$1.457\pm0.11$&$9.22\pm2.2$&$0.61\pm0.052$\\
010613......&$1.891\pm0.087$&$1.952\pm0.034$&$0.711\pm0.12$&$1.42\pm0.28$&$0.173\pm0.016$\\
010629......&$1.938\pm0.06$&$0.925\pm0.053$&$0.357\pm0.054$&$1.86\pm0.49$&$0.228\pm0.029$\\
010921......&$2.128\pm0.053$&$1.177\pm0.014$&$-0.076\pm0.028$&$1.18\pm0.01$&$0.104\pm0.001$\\
010928......&$3.293\pm0.073$&$0.841\pm0.048$&$1.814\pm0.12$&$9.23\pm1.43$&$0.406\pm0.027$\\
011019......&$1.525\pm0.178$&$1.168\pm0.089$&$-1.324\pm0.123$&$0.65\pm0.34$&$0.63\pm0.038$\\
011130......&$0.43\pm0.162$&$0.958\pm0.068$&$-2.26\pm0.06$&$0.19\pm0.05$&$0.705\pm0.038$\\
011212......&$0.471\pm0.177$&$1.476\pm0.063$&$-1.703\pm0.061$&$0.38\pm0.16$&$0.679\pm0.028$\\
020124......&$2.531\pm0.053$&$1.086\pm0.008$&$1.999\pm0.083$&$5.58\pm0.06$&$0.532\pm0.005$\\
020127......&$2.481\pm0.082$&$0.399\pm0.032$&$0.553\pm0.059$&$3.45\pm0.38$&$0.376\pm0.016$\\
020305......&$2.995\pm0.17$&$1.118\pm0.058$&$1.22\pm0.107$&$4.11\pm0.91$&$0.367\pm0.03$\\
020317......&$1.98\pm0.093$&$0.36\pm0.06$&$0.23\pm0.185$&$2.59\pm0.41$&$0.727\pm0.016$\\
020331......&$2.48\pm0.077$&$1.755\pm0.057$&$1.189\pm0.073$&$3.83\pm0.74$&$0.455\pm0.032$\\
020625......&$1.141\pm0.135$&$0.991\pm0.096$&$-1.498\pm0.126$&$0.53\pm0.29$&$0.671\pm0.035$\\
020801......&$2.054\pm0.129$&$2.339\pm0.117$&$1.241\pm0.222$&$1.68\pm0.85$&$0.507\pm0.049$\\
020812......&$2.752\pm0.182$&$0.606\pm0.103$&$1.063\pm0.121$&$4.78\pm1.39$&$0.716\pm0.035$\\
020813......&$2.773\pm0.042$&$1.588\pm0.002$&$1.982\pm0.016$&$3.8\pm0.03$&$0.169\pm0.002$\\
020819......&$1.858\pm0.079$&$1.304\pm0.067$&$-0.251\pm0.076$&$0.96\pm0.02$&$0.124\pm0.002$\\
021004......&$2.492\pm0.104$&$1.17\pm0.013$&$0.754\pm0.087$&$3.1\pm0.02$&$0.655\pm0.006$\\
021016......&$3.037\pm0.17$&$1.358\pm0.115$&$1.517\pm0.096$&$5.01\pm1.83$&$0.451\pm0.064$\\
021021......&$1.56\pm0.16$&$0.961\pm0.102$&$-0.751\pm0.143$&$1.08\pm0.3$&$0.927\pm0.009$\\
021104......&$1.84\pm0.109$&$1.002\pm0.06$&$0.097\pm0.116$&$1.25\pm0.21$&$0.77\pm0.012$\\
021112......&$2.517\pm0.148$&$-0.216\pm0.137$&$0.482\pm0.168$&$4.74\pm1.22$&$0.875\pm0.015$\\
021211......&$1.977\pm0.03$&$0.321\pm0.015$&$0.348\pm0.043$&$2.93\pm0.02$&$0.149\pm0.002$\\
030115......&$2.484\pm0.094$&$0.802\pm0.043$&$0.592\pm0.03$&$4.04\pm0.11$&$0.436\pm0.006$\\
030226......&$2.481\pm0.053$&$1.409\pm0.014$&$0.923\pm0.065$&$2.9\pm0.03$&$0.556\pm0.004$\\
030324......&$3.237\pm0.255$&$0.374\pm0.097$&$0.847\pm0.035$&$6.05\pm1.26$&$0.586\pm0.032$\\
030328......&$2.515\pm0.029$&$1.739\pm0.005$&$1.381\pm0.013$&$4.08\pm0.02$&$0.226\pm0.001$\\
030329......&$1.915\pm0.008$&$1.345\pm0.004$&$0.03\pm0.008$&$1.16\pm0.00$&$0.007\pm0.000$\\
030416......&$0.295\pm0.173$&$1.1\pm0.042$&$-2.501\pm0.139$&$0.21\pm0.03$&$0.228\pm0.01$\\
030418......&$2.309\pm0.147$&$1.558\pm0.113$&$1.175\pm0.055$&$3.19\pm1.07$&$0.906\pm0.015$\\
030429......&$2.122\pm0.069$&$0.554\pm0.054$&$0.54\pm0.049$&$3.39\pm0.02$&$0.718\pm0.004$\\
030519......&$2.409\pm0.036$&$0.841\pm0.018$&$1.21\pm0.067$&$3.74\pm0.29$&$0.075\pm0.005$\\
030528......&$1.753\pm0.043$&$1.549\pm0.02$&$0.483\pm0.035$&$1.34\pm0.01$&$0.307\pm0.003$\\
030723......&$0.793\pm0.279$&$0.795\pm0.045$&$-1.258\pm0.105$&$0.6\pm0.09$&$0.802\pm0.013$\\
030725......&$2.282\pm0.045$&$1.965\pm0.036$&$0.752\pm0.016$&$2.6\pm0.33$&$0.132\pm0.012$\\
030821......&$2.372\pm0.063$&$0.845\pm0.048$&$0.719\pm0.058$&$3\pm0.53$&$0.447\pm0.024$\\
030823......&$1.717\pm0.071$&$1.439\pm0.043$&$0.101\pm0.054$&$1.14\pm0.23$&$0.53\pm0.022$\\
030824......&$0.74\pm0.17$&$0.915\pm0.041$&$-1.659\pm0.089$&$0.29\pm0.1$&$0.551\pm0.03$\\
030913......&$2.996\pm0.147$&$-0.012\pm0.137$&$1.023\pm0.189$&$8.19\pm2.21$&$0.772\pm0.032$\\
031026......&$3.595\pm0.234$&$0.947\pm0.093$&$2.178\pm0.136$&$6.89\pm1.69$&$0.969\pm0.004$\\
031109A......&$2.564\pm0.044$&$1.47\pm0.024$&$1.02\pm0.025$&$2.93\pm0.38$&$0.129\pm0.011$\\
031109B......&$2.231\pm0.174$&$2.145\pm0.129$&$0.664\pm0.099$&$3.09\pm1.18$&$0.94\pm0.01$\\
031111A......&$3.116\pm0.045$&$0.402\pm0.037$&$1.288\pm0.053$&$10.51\pm1.17$&$0.182\pm0.018$\\
031111B......&$0.752\pm0.272$&$1.359\pm0.027$&$-1.808\pm0.14$&$0.19\pm0.05$&$1$\\
031203......&$2.683\pm0.074$&$0.517\pm0.057$&$1.661\pm0.057$&$6.63\pm1.31$&$0.26\pm0.034$\\
031220......&$2.204\pm0.19$&$0.585\pm0.117$&$0.013\pm0.062$&$2.08\pm0.82$&$0.64\pm0.048$\\
040319......&$2.183\pm0.129$&$0.37\pm0.124$&$0.5\pm0.13$&$2.9\pm1.31$&$0.421\pm0.062$\\
040423......&$1.845\pm0.085$&$1.314\pm0.078$&$0.607\pm0.102$&$1.8\pm0.58$&$0.507\pm0.041$\\
040425......&$3.001\pm0.064$&$1.634\pm0.038$&$1.307\pm0.07$&$2.73\pm0.35$&$0.739\pm0.009$\\
040511......&$2.421\pm0.062$&$1.22\pm0.05$&$1.255\pm0.076$&$3.46\pm0.63$&$0.366\pm0.03$\\
040701......&$0.405\pm0.146$&$0.955\pm0.122$&$-2.096\pm0.074$&$0.3\pm0.01$&$0.384\pm0.013$\\
040709......&$2.15\pm0.051$&$1.618\pm0.041$&$0.898\pm0.035$&$2.49\pm0.5$&$0.182\pm0.021$\\
040802......&$2.43\pm0.077$&$0.033\pm0.062$&$0.955\pm0.115$&$3.74\pm0.79$&$0.358\pm0.036$\\
040825A......&$2.056\pm0.133$&$1.287\pm0.068$&$-0.088\pm0.069$&$1.17\pm0.35$&$0.772\pm0.02$\\
040825B......&$1.942\pm0.123$&$0.701\pm0.098$&$0.698\pm0.039$&$2.49\pm0.52$&$0.838\pm0.014$\\
040912A......&$1.247\pm0.095$&$0.851\pm0.064$&$-0.782\pm0.24$&$0.61\pm0.36$&$0.162\pm0.043$\\
040912B......&$1.62\pm0.194$&$1.68\pm0.015$&$0.327\pm0.131$&$1.82\pm0.01$&$0.754\pm0.007$\\
040916......&$0.465\pm0.163$&$2.401\pm0.052$&$-1.329\pm0.052$&$0.42\pm0.18$&$0.856\pm0.028$\\
040924......&$1.88\pm0.014$&$0.259\pm0.006$&$0.292\pm0.029$&$2.5\pm0.02$&$0.129\pm0.001$\\
041004......&$1.938\pm0.059$&$1.504\pm0.037$&$0.632\pm0.235$&$1.13\pm0.25$&$0.213\pm0.019$\\
041006......&$1.915\pm0.016$&$1.11\pm0.004$&$0.219\pm0.013$&$2.37\pm0.02$&$0.092\pm0.001$\\
041016......&$3.263\pm0.288$&$0.689\pm0.096$&$1.64\pm0.243$&$4.95\pm1.35$&$0.7\pm0.031$\\
041127......&$1.809\pm0.091$&$1.417\pm0.067$&$0.458\pm0.08$&$1.53\pm0.49$&$0.341\pm0.038$\\
041211......&$2.804\pm0.188$&$1.42\pm0.054$&$1.496\pm0.263$&$8.66\pm1.5$&$0.389\pm0.029$\\
050123......&$2.013\pm0.092$&$0.765\pm0.085$&$0.434\pm0.142$&$2.73\pm0.86$&$0.395\pm0.045$\\
050209......&$3.311\pm0.274$&$0.669\pm0.111$&$1.2\pm0.212$&$4.24\pm1.36$&$0.663\pm0.043$\\
050408......&$1.778\pm0.083$&$1.103\pm0.005$&$0.537\pm0.007$&$2.68\pm0.01$&$0.261\pm0.002$\\
050509......&$1.104\pm0.345$&$1.12\pm0.095$&$-0.836\pm0.125$&$0.64\pm0.28$&$0.807\pm0.026$\\
050729......&$2.455\pm0.084$&$0.172\pm0.059$&$1.515\pm0.214$&$4.87\pm0.86$&$0.447\pm0.026$\\
050807......&$2.25\pm0.224$&$0.781\pm0.117$&$0.15\pm0.117$&$1.14\pm0.67$&$0.411\pm0.066$\\
050922......&$2.656\pm0.073$&$0.285\pm0.062$&$0.742\pm0.132$&$4.59\pm0.08$&$0.378\pm0.007$\\
051021......&$2.386\pm0.067$&$1.184\pm0.053$&$0.685\pm0.061$&$3.04\pm0.65$&$0.289\pm0.028$\\
051022......&$2.585\pm0.022$&$1.995\pm0.012$&$1.356\pm0.014$&$3.71\pm0.05$&$0.065\pm0.001$\\
051028......&$3.103\pm0.152$&$0.509\pm0.1$&$1.273\pm0.125$&$6.72\pm2.05$&$0.522\pm0.049$\\
051211......&$2.862\pm0.101$&$-0.076\pm0.085$&$2.24\pm0.377$&$7.71\pm1.85$&$0.652\pm0.035$\\
060115......&$2.397\pm0.056$&$0.91\pm0.025$&$1.096\pm0.065$&$4.63\pm0.46$&$0.203\pm0.014$\\
060121......&$2.526\pm0.043$&$-0.045\pm0.033$&$1.499\pm0.069$&$5.37\pm0.62$&$0.253\pm0.018$\\
\hline
\end{longtable}
}
}
\end{document}